\documentclass[fleqn,usenatbib]{mnras}

\usepackage{newtxtext,newtxmath}

\usepackage[T1]{fontenc}

\DeclareRobustCommand{\VAN}[3]{#2}
\let\VANthebibliography\thebibliography
\def\thebibliography{\DeclareRobustCommand{\VAN}[3]{##3}\VANthebibliography}

\usepackage{graphicx}	
\usepackage{amsmath}	
\usepackage{float}             
\usepackage{threeparttable}    
\usepackage{scalerel}
\usepackage{tikz}
\usetikzlibrary{svg.path}

\definecolor{orcidlogocol}{HTML}{A6CE39}
\tikzset{
  orcidlogo/.pic={
    \fill[orcidlogocol] svg{M256,128c0,70.7-57.3,128-128,128C57.3,256,0,198.7,0,128C0,57.3,57.3,0,128,0C198.7,0,256,57.3,256,128z};
    \fill[white] svg{M86.3,186.2H70.9V79.1h15.4v48.4V186.2z}
                 svg{M108.9,79.1h41.6c39.6,0,57,28.3,57,53.6c0,27.5-21.5,53.6-56.8,53.6h-41.8V79.1z M124.3,172.4h24.5c34.9,0,42.9-26.5,42.9-39.7c0-21.5-13.7-39.7-43.7-39.7h-23.7V172.4z}
                 svg{M88.7,56.8c0,5.5-4.5,10.1-10.1,10.1c-5.6,0-10.1-4.6-10.1-10.1c0-5.6,4.5-10.1,10.1-10.1C84.2,46.7,88.7,51.3,88.7,56.8z};
  }
}
\newcommand\orcidicon[1]{\href{https://orcid.org/#1}{\mbox{\scalerel*{
\begin{tikzpicture}[yscale=-1,transform shape]
\pic{orcidlogo};
\end{tikzpicture}
}{|}}}}

\definecolor{pgcolor}{HTML}{7788FF}

\title[Inferring the SMF at z > 6 using Local Group galaxies]{Testing the near-far connection with FIRE simulations: inferring the stellar mass function of the proto-Local Group at z > 6 using the fossil record of present-day galaxies}

\author[Pratik J. Gandhi et al.]{
Pratik J. Gandhi\orcidicon{0000-0003-0965-605X}$^{1}$\thanks{E-mail: pjgandhi@ucdavis.edu; gandhipratik1995@gmail.com}\thanks{Frontera Computational Science Fellow},
Andrew Wetzel\orcidicon{0000-0003-0603-8942}$^{1}$,
Michael Boylan-Kolchin\orcidicon{0000-0002-9604-343X}$^{2}$,
Robyn E. Sanderson\orcidicon{0000-0003-3939-3297}$^{3,4}$,
\newauthor
\hspace{0.25mm} Alessandro Savino\orcidicon{0000-0002-1445-4877}$^{5}$, Daniel R. Weisz\orcidicon{0000-0002-6442-6030}$^{5}$, Erik J. Tollerud\orcidicon{0000-0002-9599-310X}$^{6}$, Guochao Sun\orcidicon{0000-0003-4070-497X}$^7$, and 
\newauthor
\hspace{0.25mm} Claude-André Faucher-Giguère\orcidicon{0000-0002-4900-6628}$^{7}$
\\
$^{1}$Department of Physics and Astronomy, University of California - Davis, One Shields Avenue, Davis, CA 95616, USA\\
$^{2}$Department of Astronomy, The University of Texas at Austin, 2515 Speedway, Austin, TX 78712, USA\\
$^{3}$Department of Physics and Astronomy, University of Pennsylvania, 209 South 33rd Street, Philadelphia, PA 19104, USA\\
$^{4}$Center for Computational Astrophysics, Flatiron Institute, 162 Fifth Avenue, New York, NY 10010, USA\\
$^{5}$Department of Astronomy, University of California, Berkeley, Berkeley, CA, 94720, USA\\
$^{6}$Space Telescope Science Institute, 3700 San Martin Drive, Baltimore, MD 21218, USA\\
$^{7}$CIERA and Department of Physics and Astronomy, Northwestern University, 1800 Sherman Avenue, Evanston, IL 60201, USA
}

\date{Accepted XXX. Received YYY; in original form ZZZ}

\pubyear{2015}

\begin{document}
\label{firstpage}
\pagerange{\pageref{firstpage}--\pageref{lastpage}}
\maketitle

\begin{abstract}
The shape of the low-mass (faint) end of the galaxy stellar mass function (SMF) or ultraviolet luminosity function (UVLF) at $z \gtrsim 6$ is an open question for understanding which galaxies primarily drove cosmic reionisation. Resolved photometry of Local Group low-mass galaxies allows us to reconstruct their star formation histories, stellar masses, and UV luminosities at early times, and this fossil record provides a powerful `near-far' technique for studying the reionisation-era SMF/UVLF, probing orders of magnitude lower in mass than direct HST/JWST observations. Using $882$ low-mass ($M_{\rm star}\lesssim10^{9}\,\rm{M_\odot}$) galaxies across $11$ Milky Way- and Local Group-analogue environments from the FIRE-2 cosmological baryonic zoom-in simulations, we characterise their progenitors at $z=6-9$, the mergers/disruption of those progenitors over time, and how well their present-day fossil record traces the high-redshift SMF. A present-day galaxy with $M_{\rm star}\sim10^5\,\rm{M_\odot}$ ($\sim10^9\,\rm{M_\odot}$) had $\approx1$ ($\approx30$) progenitors at $z\approx7$, and its main progenitor comprised $\approx100\%$ ($\approx50\%$) of the total stellar mass of all its progenitors at $z\approx7$. We show that although only $\sim 15\%$ of the early population of low-mass galaxies survives to present day, the fossil record of surviving Local Group galaxies accurately traces the low-mass slope of the SMF at $z \sim 6 - 9$. We find no obvious mass dependence to the mergers and accretion, and show that applying this reconstruction technique to just the low-mass galaxies at $z = 0$ and not the MW/M31 hosts correctly recovers the slope of the SMF down to $M_{\rm star} \sim 10^{4.5}\,\rm{M_{\odot}}$ at $z \gtrsim 6$ . Thus, we validate the `near-far' approach as an unbiased tool for probing low-mass reionisation-era galaxies.
\end{abstract}




\section{Introduction}
\label{sec:1-intro}

\subsection{Motivating questions}
\label{sec:1.x-motivation}

The Epoch of Reionisation (EoR), during which the hydrogen gas in the intergalactic medium (IGM) went from being neutral to ionised, was one of the most important phase transitions in the history of the Universe. The current consensus is that energetic radiation from the first star-forming galaxies predominantly drove cosmic reionisation, but major questions remain about the nature of the galaxies that contributed most to the overall ionising photon budget. Most models of reionisation argue that low- to intermediate-mass galaxies contributed more total ionising photons than brighter, more massive galaxies, because of their larger numbers and higher ionising photon escape fractions \citep[e.g.,][]{kuhlen-12, wise-14, robertson-15, ma-18-b} \citep[though see e.g.,][for an example of analyses that favour more massive galaxies for driving the bulk of reionisation]{naidu-20}. To address this question of whether lower mass, fainter galaxies did indeed drive a majority of the reionisation process, we need to understand the shape of the galaxy stellar mass function (SMF) and rest-frame ultra-violet luminosity function (UVLF) during the EoR at $z \gtrsim 6$. \textit{A key specific question is: what is the slope of the galaxy SMF/UVLF at the low-mass/faint end at $z \gtrsim 6$?}

Direct HST observations have provided strong constraints on the UVLF at $z \sim 6 - 9$ for galaxies as faint as $M_{\rm UV} \sim -17$, and studies that leveraged the power of gravitational lensing have provide information about systems that are $\approx 2-3$ orders of magnitude fainter \citep[][amongst others]{finkelstein-15, bouwens-17, livermore-17, atek-18}. More recently, JWST is returning exquisite measurements of high-redshift galaxies, with surveys like CEERS, JADES, and NGDEEP poised to push the frontier of the reionisation-era galaxy SMF/UVLF down to fainter magnitudes \citep[see][]{leung-23, navarro-carrera-23, perez-gonzalez-23}. However, even deep HST/JWST imaging is likely unable to constrain the faintest (and most numerous) of these high-redshift galaxies during the EoR -- as faint as $M_{\rm UV} \sim -3$ to $-6$ at $z \gtrsim 6$ -- making it difficult to study these likely drivers of reionisation \citep[e.g.,][]{boylan-kolchin-16, weisz-17}. Direct observations become substantially more uncertain in the regime where results come exclusively from gravitational lensing ($M_{\rm UV} \gtrsim -15$), due to systematic uncertainties in lens models and magnification maps, and sometimes due to contamination from the lensing cluster itself \citep[e.g.,][]{bouwens-17, atek-18}, posing another limitation on directly measuring the low-mass/faint end slope of the SMF/UVLF during the EoR.

Beyond their potentially significant contributions to reionisation, we also know that faint, low-mass galaxies constitute a majority of the galaxy population in the universe at all redshifts. Low-mass galaxies are also excellent probes of various astrophysical phenomena like stellar feedback, quenching due to reionisation, enrichment from PopIII stars, and the nature of dark matter. However, significant questions remain about their nature and population demographics at early cosmic times because of the difficulty of directly observing such faint galaxies. This provides further motivation for studying low-mass galaxies during the EoR, to build a holistic theory of galaxy formation across a wide range of masses.


\subsection{A `near-far' solution}
\label{sec:1.x-near-far-approach}

A novel alternative to direct observations for studying low-mass/faint galaxies at $z \gtrsim 6$ has emerged in recent years \citep{weisz-14c, boylan-kolchin-15, boylan-kolchin-16, weisz-17}. This `near-far' technique bridges the gap between near-field HST/JWST observations of low-mass galaxies in the Local Group and faint galaxies in the early universe. It leverages resolved photometry and colour-magnitude diagram (CMD) modelling of low-mass galaxies ($M_{\rm star} \lesssim 10^9 \, \rm{M_\odot}$) in the Local Group (LG) to reconstruct their stellar masses ($M_{\rm star}$), star formation histories (SFHs), and rest-frame UV luminosities at $z \gtrsim 6$ \citep[for example][]{brown-14, weisz-14b, weisz-14a, geha-15, skillman-17, savino-23}. By using the stellar fossil record of low-mass galaxies in the LG, one can infer the low-mass (faint) end slope of the SMF/UVLF during reionisation, much deeper than direct observations at $z \gtrsim 6$ can.

\textit{Crucially, the LG is the only place in the universe where we can observe and reconstruct SFHs for galaxies as low in mass as $M_{\rm star}(z = 0) \sim 10^3 \, \rm{M_\odot}$.} Because these low-mass galaxies are likely the descendants of the faintest, lowest-mass galaxies during reionisation, the LG remains the only place in which this kind of complementary technique to direct observations for studying the earliest galaxies is possible. Additionally, even for galaxies that are faint or low-mass at high redshift but still observable by HST/JWST, determining their stellar masses and UV luminosities is quite tricky due to large uncertainties in techniques used for modelling their star formation histories (for example). The `near-far' technique therefore probes a regime of galaxy formation that is beyond the scope of current and even upcoming observations with the largest ground- and space-based observatories. Even in the worst-case scenario of how low in mass (or faint) it can probe, it still provides a strong complementary approach to studying the galaxy SMF/UVLF at high redshift against which we can compare results from direct observations. Hence, rigorously stress-testing the near-far approach to quantify all sources of uncertainty is essential, to understand just how accurate the inference of the high-redshift low-mass galaxy population using the fossil record of present-day galaxies is.

\subsection{Key uncertainties in the near-far approach}
\label{sec:1.x-near-far-uncertainties}

Previous studies like \citet{weisz-14c} have inferred a faint-end UVLF slope at $z \gtrsim 6$ in this manner. However, a few critical questions remain about the efficacy of this near-far reconstruction technique. The first is related to the accuracy of the CMD-based method for reconstructing SFHs and UV luminosities, which includes uncertainties in both measurements and stellar evolution models. There is also a preponderance of evidence (from both LG and high-redshift observations) for the bursty nature of the SFHs of low-mass galaxies at early cosmic times \citep[see][]{mcquinn-10a, mcquinn-10b, weisz-12, sparre-17, ma-18-b, emami-21, flores-velazquez-21, furlanetto-22, dressler-23, looser-23, pallottini-23, shen-23, sun-23}, and the CMD-based reconstruction method can only probe down to a limiting baseline when it comes to short duty cycles of fluctuations in UV luminosity arising from bursty star formation. This further adds uncertainty to our ability to recover the slope of the reionisation-era UVLF using the fossil record of LG galaxies. \textit{In this paper, we focus solely on testing how accurately the near-far technique infers the low-mass end slope of the SMF at $z \sim 6 - 9$, and not the UVLF. In an upcoming Gandhi et al., in prep. paper,  we will use synthetic observations of low-mass galaxies in the FIRE simulations to do so.}

If one assumes that the overall galaxy distribution in the proto-LG is representative of the global galaxy population in the universe \citep[][showed how this should be a reasonable assumption for progenitors of present-day ultra-faint and classical dwarf spheroidals]{boylan-kolchin-16}, then a second key question about the near-far technique is how well the fossil record of \textit{surviving} low-mass galaxies in the LG at $z = 0$ represents the true low-mass galaxy population in the proto-LG at $z \gtrsim 6$. This is uncertain because of two effects: (a) mergers of low-mass galaxies with each other over cosmic time, and (b) the accretion/disruption of low-mass galaxies as they fall into the central Milky Way (MW) or Andromeda (M31) galaxy. Because reconstructing SFHs for low-mass galaxies at $z = 0$ only tells us what their total progenitor stellar mass was at $z \gtrsim 6$ but not how many progenitor galaxies that mass was distributed amongst. The fossil record contains virtually no information about the mergers and disruptions that led to the low-mass galaxy population we see at present day. Thus, mergers between low-mass galaxies in the proto-LG as well as low-mass galaxies accreting onto the central MW/M31 could potentially bias the inference of the slope of SMF at high redshift. Since the nature of mergers and disruptions of low-mass galaxies depends on baryonic physics, we need to consider fully baryonic simulations to model them instead of just dark matter-only simulations or other less comprehensive techniques. \textit{In this paper, we use the FIRE-2 simulations of galaxy formation to provide a theoretical characterization of how mergers, accretion, and disruption of low-mass galaxies in the proto-Local Group impact the near-far reconstruction technique for studying low-mass galaxies during the EoR.
}

\section{Methods}
\label{sec:2-methods}

\subsection{FIRE-2 simulations}
\label{sec:2.x-fire-sims} 

We use the Latte \citep[introduced in][]{wetzel-16} and ELVIS on FIRE \citep[introduced in][]{garrison-kimmel-19a} suites from the FIRE-2 cosmological baryonic zoom-in simulations \citep{hopkins-18a} of the Feedback In Realistic Environments (FIRE) project\footnote{\url{https://fire.northwestern.edu}}.
These simulations model $7$ isolated MW-analogue and $3$ paired LG-analogue galaxies along with their surrounding low-mass galaxies.
The $3$ pairs of ELVIS simulations (ELVIS) have a mass resolution of $m_{\rm baryon,ini} = 3500 -4200 \, \rm{M_\odot}$ ($m_{\rm dm} \approx 2 \times 10^4 \, \rm{M_\odot}$), and the other $7$ isolated hosts (Latte) have $m_{\rm baryon,ini} = 7100 \, \rm{M_\odot}$ ($m_{\rm dm} = 3.5 \times 10^4 \, \rm{M_\odot}$). 
The host halos have total masses $M_{\rm 200m} \approx 1 - 2 \times 10^{12} \, \rm{M_\odot}$, which are within observational uncertainties of the MW's properties\footnote{`200m' indicates a measurement relative to 200 times the mean matter density of the Universe.}. The central galaxy stellar masses are $M_{\rm star} \approx 10^{10 - 11} \, \rm{M_\odot}$.
Crucially, these simulations reproduce the stellar mass functions, radial distance distributions, and star-formation histories of low-mass galaxies in the LG \citep{wetzel-16, garrison-kimmel-19a, garrison-kimmel-19b, samuel-20, samuel-21}, providing a reliable sample of $882$ low-mass galaxies at $z = 0$ to work with.

Table~\ref{tab:FIRE-2-sims} lists the FIRE-2 simulations that we use along with their properties at $z = 0$, including: baryonic resolution, central host(s) stellar mass, $M_{\rm 200m}$, $R_{\rm 200m}$, and the total number of low-mass galaxies with $10^{4.5}\,\rm{M_{\odot}} \leq M_{\rm star} \leq 10^{9} \rm{M_\odot}$ out to $2$ Mpc from the isolated hosts or $2$ Mpc from the geometric centre of the paired hosts.

The FIRE-2 simulations are run using \textsc{Gizmo}, a Lagrangian Meshless Finite Mass (MFM) hydrodynamics code \citep{hopkins-15}. Each simulation includes an implementation of fluid dynamics, star formation, and stellar feedback based on the FIRE-2 numerical prescription. FIRE-2 models the dense, multi-phase interstellar medium (ISM) in galaxies and incorporates physically motivated, metallicity-dependent radiative heating and cooling processes for gas. These include free-free, photoionisation and recombination, Compton, photo-electric and dust collisional, cosmic ray, molecular, metal-line, and fine structure processes. They account for $11$ element species (H, He, C, N, O, Ne, Mg, Si, S, Ca, Fe) across a temperature range of $10 - 10^{10} \rm{K}$. The simulations also include the subgrid diffusion and mixing of these elements in gas via turbulence \citep[see][for further details]{escala-18, hopkins-18a}. Additionally, the FIRE-2 simulations model the global effects of cosmic reionisation on gas using a spatially uniform, redshift-dependent meta-galactic UV/X-ray background based on an update to \citet{faucher-giguere-09}\footnote{see \url{https://galaxies.northwestern.edu/uvb-fg09} for details on this December 2011 update, which was designed to reionise by $z\sim10$, as was preferred by empirical constraints at the time these simulations were run.}, which we discuss further in Section~\ref{sec:4.x-uvbackground}.

Star particles in the FIRE-2 model form out of gas that is self-gravitating, Jeans-unstable, cold ($T < 10^{4} \, \rm{K}$), dense ($n > 10^3 \, \rm{cm}^{-3}$), and molecular \citep[following][]{krumholz-11}. Each star particle represents a single stellar population, assuming a \citet{kroupa-01} stellar initial mass function. During formation, star particles also inherit the mass and elemental abundances of their respective progenitor gas cells. In FIRE-2, star particles evolve along standard stellar population tracks from STARBURST99 v7.0 \citep{leitherer-99}. We also include the following time-resolved stellar feedback processes: core-collapse and white-dwarf (Type Ia) supernovae, continuous mass loss, radiation pressure, photoionisation, and photoelectric heating. FIRE-2 uses rates for core-collapse supernovae from STARBURST99 \citep{leitherer-99} and nucleosynthetic yields from \citet{nomoto-06}. Stellar wind yields, sourced primarily from O, B, and AGB stars, are from a combination of models from \citet{van-den-hoek-97}, \citet{marigo-01}, and \citet{izzard-04}, synthesized in \citet{wiersma-09}. For white-dwarf supernovae, FIRE-2 uses rates from \citet{mannucci-06} and nucleosynthetic yields from \citet{iwamoto-99}. For a more detailed discussion of the implementation of supernova feedback, see \citet{hopkins-18b}.

We generate cosmological zoom-in initial conditions for each simulation at $z \approx 99$ using the \textsc{MUSIC} code \citep{hahn-11}. These initial conditions are embedded within periodic cosmological boxes of length $70$ to $172$ Mpc. We save 600 snapshots per simulation from $z \approx 99$ to $z = 0$, with an average spacing of $\lesssim 25$ Myr. For all simulations we assume flat $\Lambda$CDM cosmology, using parameters broadly consistent with \citet{planck-20}: $h = 0.68 - 0.71$, $\Omega_\Lambda = 0.69 - 0.734$, $\Omega_{\rm m} = 0.266 - 0.31$, $\Omega_{\rm b} = 0.0455 - 0.048$, $\sigma_8 = 0.801 - 0.82$, and $n_{\rm s} = 0.961 - 0.97$.

\begin{table*}
\centering
\caption{
\textbf{
Properties at $z = 0$ of the FIRE-2 simulations in our sample.} 
`MW-analogues' are simulations of isolated MW-mass systems with surrounding low-mass galaxies, while `LG-analogues' contain a paired MW and M31 with surrounding low-mass galaxies. We list the initial masses of star particles and gas cells under `Baryonic Resolution’. We measure stellar masses for the central host galaxies within a spherical volume of radius $15$ kpc from the galaxies' centres, while $M_{\rm 200m}$ is the total mass within their virial radius, $R_{\rm 200m}$. For the MW-analogues, we consider all low-mass galaxies within $2$ Mpc from the host, while for the LG-analogues we consider all low-mass galaxies within $2$ Mpc from the geometric centre of the two hosts.
}
\addtolength{\tabcolsep}{1pt}
\begin{tabular}{|c|c|c|c|c|c|c|c|}
\hline
\hline
Simulation name & Type of & Baryonic & $M_{\rm star}$ of & $M_{200 \rm m}$ of & $R_{200 \rm m}$ of & Number of low- & Reference$^{\dagger}$\\
& simulation & Resolution & host(s) & host(s) & host(s) & mass galaxies & \\
& & [$\rm{M}_\odot$] & [$\times 10^{10} \; \rm{M}_\odot$] & [$\times 10^{12} \; \rm{M}_\odot$] & [kpc] & ($10^{4.5} \leq M_{*} \leq 10^9 \rm{M}_\odot$) & \\
& & & at $z = 0$ & at $z = 0$ & at $z = 0$ & at $z = 0$ &\\
\hline
m12i & MW-analogue & $7100$ & $6.3$ & $1.1$ & $328$ & $34$ & A\\
m12f & MW-analogue & $7100$ & $8.5$ & $1.6$ & $368$ & $58$ & B\\
m12m & MW-analogue & $7100$ & $12.0$ & $1.5$ & $360$ & $80$ & C\\
m12b & MW-analogue & $7100$ & $8.2$ & $1.3$ & $350$ & $53$ & D\\
m12c & MW-analogue & $7100$ & $6.1$ & $1.3$ & $342$ & $85$ & D\\
m12r & MW-analogue & $7100$ & $1.8$ & $0.96$ & $304$ & $50$ & E\\
m12w & MW-analogue & $7100$ & $5.5$ & $0.91$ & $301$ & $86$ & E\\
Romeo \& Juliet & LG-analogue & $3500$ & $7.3$   \&   $3.7$ & $1.1$   \&   $0.92$ & $317$ \& $302$ & $152$ & D\\
Romulus \& Remus & LG-analogue & $4000$ & $10.0$   \&   $4.9$ & $1.7$   \&   $1.0$ & $375$ \& $320$ & $141$ & F\\
Thelma \& Louise & LG-analogue & $4000$ & $7.7$   \&   $2.7$ & $1.1$   \&   $0.94$ & $332$ \& $310$  & $143$ & D\\
\hline
\hline
\end{tabular}
\addtolength{\tabcolsep}{-1pt}
\label{tab:FIRE-2-sims}
\begin{tablenotes}
\item $^{\dagger}$Simulation first introduced at this resolution in: A: \citet{wetzel-16}, B: \citet{garrison-kimmel-17}, C: \citet{hopkins-18a}, D: \citet{garrison-kimmel-19a}, E: \citet{samuel-20}, and F: \citet{garrison-kimmel-19b}.
\end{tablenotes}
\end{table*}

\subsection{Catalogues of halos and galaxies}
\label{sec:2.x-haloes}

We identify dark matter(DM) haloes and sub-haloes using the \textsc{ROCKSTAR} 6D-phase space finder \citep{behroozi-13a}, according to the radius that encloses 200 times the mean matter density ($R_{\rm 200m}$), and we keep those haloes and sub-haloes that have bound mass fractions $> 0.4$ and at least $30$ dark-matter particles each. We generate a halo catalogue at each of the $600$ snapshots for each simulation, using only DM particles.

We then assign star particles to each halo and sub-halo in post-processing as follows \citep[adapted from the method in][]{necib-19, samuel-20}. The assignment varies slightly for $z = 0$ versus $z \geq 6$, as described below. At $z = 0$, given each (sub)halo’s radius, $R_{\rm 200m}$, and $v_{\rm circ, max}$ from \textsc{ROCKSTAR}, we first identify all star particles whose position is within $0.8\,R_{\rm 200m}$ (out to a maximum radius of $30$ kpc) and whose velocity is within $2 \, v_{\rm circ, max}$ of each (sub)halo’s centre-of-mass velocity. We then keep star particles whose (a) whose positions are within $1.5 \, R_{90}$ (the radius that encloses $90$ per cent of the mass of member star particles) of both the centre-of-mass position of member stars and the halo centre (thus ensuring the galaxy centre is coincident with the halo centre) and (b) velocities are within $2 \sigma_{\rm vel, star}$ of the centre-of-mass velocity of member stars. We then iteratively repeat (1) and (2) until $M_{\rm star}$, the sum of the masses of all member star particles, converges to within $1$ per cent. At $z = 0$, we keep all (sub)haloes with at least $6$ star particles and average stellar density $> 300 \, \rm{M}_\odot \rm{kpc}^{-3}$. These criteria ensure that we distinguish true galaxies from transient alignments between subhaloes and stars in the stellar halo of a more massive galaxy such as the central MW/M31-mass host.

At $z \geq 6$, we use a similar approach but with different numerical parameters. The radius for including star particles is larger ($1.0 \, R_{200\rm m}$ instead of $0.8 \, R_{200\rm m}$), and we do not apply any cuts on velocity relative to the (sub)halo's centre-of-mass velocity. We use these parameters at high redshifts to improve the overall completeness of star particles assigned to a halo.
Additionally, we keep all haloes with at least $2$ star particles, given that we mostly only analyse galaxies at high redshift that are the progenitors of well-resolved galaxies at $z = 0$.

\subsection{Selecting low-mass galaxies at {$z = 0$} and identifying progenitors at high redshift}
\label{sec:2.x-selecting-dwarfs}

We use the same selection of low-mass galaxies at $z = 0$ throughout our analysis: all galaxies with $M_{\rm star} \geq 10^{4.5} \, \rm{M_\odot}$, out to $2$ Mpc from the center of each isolated MW-analogue or from the geometric centre of each LG-analogue pair. We do not include the central MW/M31-mass galaxies in our sample at $z = 0$. Across our suite of simulations, this is the typical maximum distance without significant contamination from low-resolution DM particles near the outskirts of each zoom-in region. With JWST making it possible to resolve stellar populations farther out in the Local Group than HST and thus extending the reach of the near-far reconstruction technique, we choose a sample of low-mass galaxies to go as far out as possible. In an upcoming Gandhi et al., paper, we will explore how our results vary with selection distance of galaxies at $z = 0$.

We then track all the star particles in our selected galaxies at $z = 0$ back to $z = 6 - 9$ to identify their progenitor galaxies in the simulation volume that are (a) not contaminated by low-resolution DM particles, (b) have at least $2$ star particles, and (c) have at least $1$ star particle ending up in our galaxy sample at $z = 0$. Therefore, in most cases the galaxy sample at $z \geq 6$ does not include galaxies that eventually formed the central MW/M31-mass galaxy; however, in some figures we do include them (we note such cases; in particular Figures~\ref{fig:mtot-populations-vs-z}, \ref{fig:smf-z0-z7}, \ref{fig:different-smfs-compared-to-all-at-z7}, and \ref{fig:slope-comparison}), mostly when we make comparisons to the total galaxy population in the entire progenitor systems of the present-day MW/LG-analogue systems. Additionally, the `main progenitor' is the one with the highest stellar mass out of all the progenitor galaxies at a given redshift that contribute at least $4$ star particles to the specific low-mass galaxy at $z = 0$. This additional criterion of $4$ star particles is used to avoid spurious cases in which a high-redshift galaxy that is actually a progenitor of the central MW/M31-like host, gets tagged as a progenitor of a low-mass galaxy because for some reason it contributed $1$ star particle to it (due to either physical or numerical effects). 

Of all our low-mass galaxies at $z = 0$, a small number ($\approx 7$ per cent) have stellar mass traceable back to $z \geq 6$ but are not associated with any progenitor halo(es). These are mainly the lowest-mass galaxies whose stellar mass at $z = 0$ is close to the resolution limit of our simulations, and because we only consider high-redshift galaxies with at least $2$ star particles, their progenitor haloes likely contain only $1$ star particle and thus do not end up in our catalogue. For these cases, we assert that the galaxy has $1$ progenitor at that redshift, and we assign all the stellar mass that we track back to that single progenitor.

\section{Results}
\label{sec:3-results}

\textit{A reminder that in this paper we focus solely on testing the accuracy of inferring the slope of the reionisation-era SMF (and not the UVLF) using the near-far reconstruction method.} We show most of our key results in Sections~\ref{sec:3.x-mstar-at-z7}, \ref{sec:3.x-smf-z0-vs-z7}, \ref{sec:3.x-prog-at-z7}, and \ref{sec:3.x-all-z7-gal-vs-reconstruction} at our `fiducial' redshift of $z = 7$. This is because the mid-point of the Epoch of Reionisation -- when the cosmic neutral hydrogen fraction is $\approx 0.5$ -- is empirically estimated to be between $z\sim7$ and $z\sim8$ \citep{planck-20}. Towards the end of Section~\ref{sec:3.x-all-z7-gal-vs-reconstruction}, we discuss how our key results vary (or not) from $z = 6 - 9$, to get a better sense of the range over a larger period during the EoR.

\subsection{Total stellar mass budget at $z > 6$} 
\label{sec:3.x-mstar-at-z7}

\begin{figure*}
\centering
\begin{tabular}{c c}
\includegraphics[width = 0.45 \linewidth]{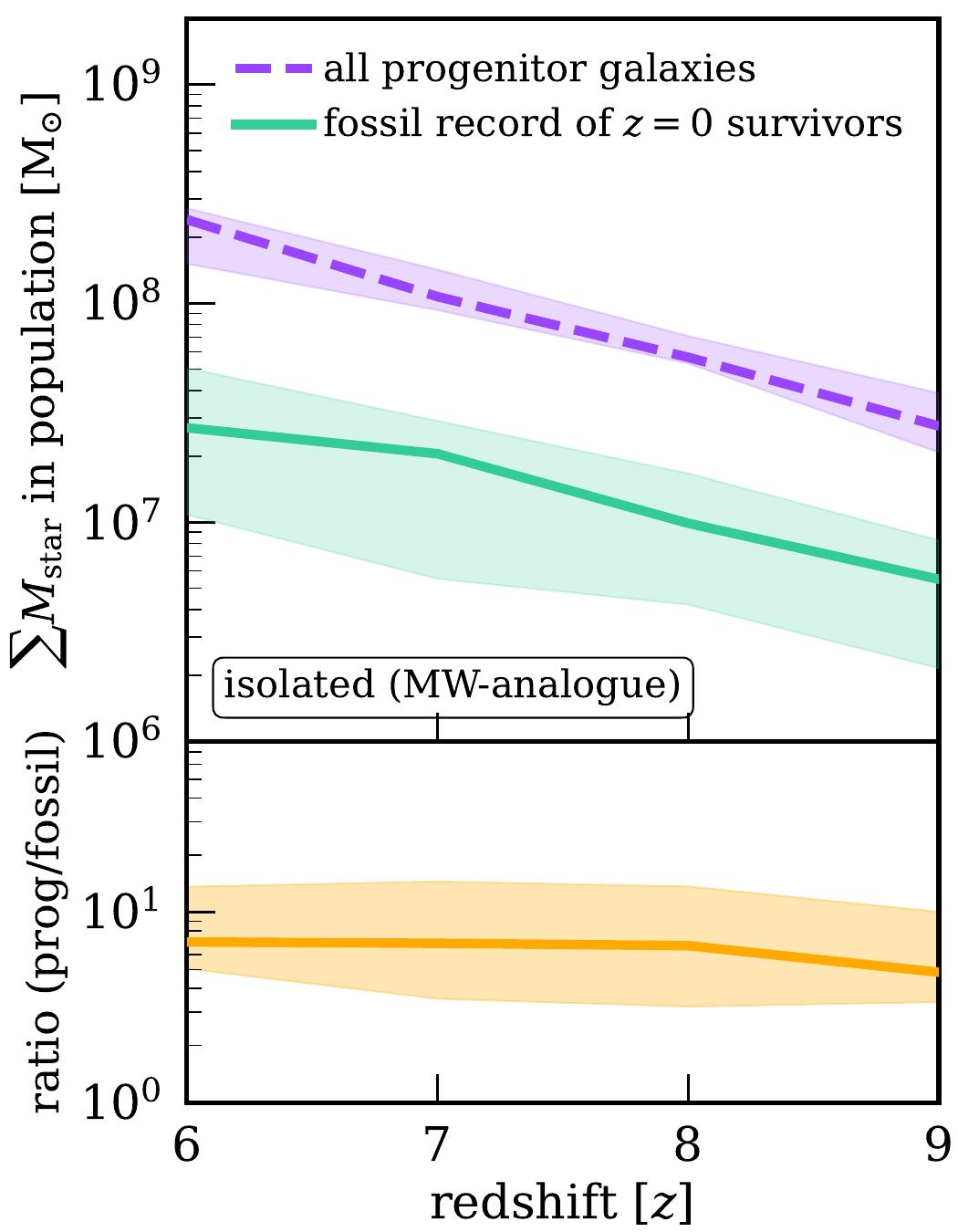}&
\includegraphics[width = 0.45 \linewidth]{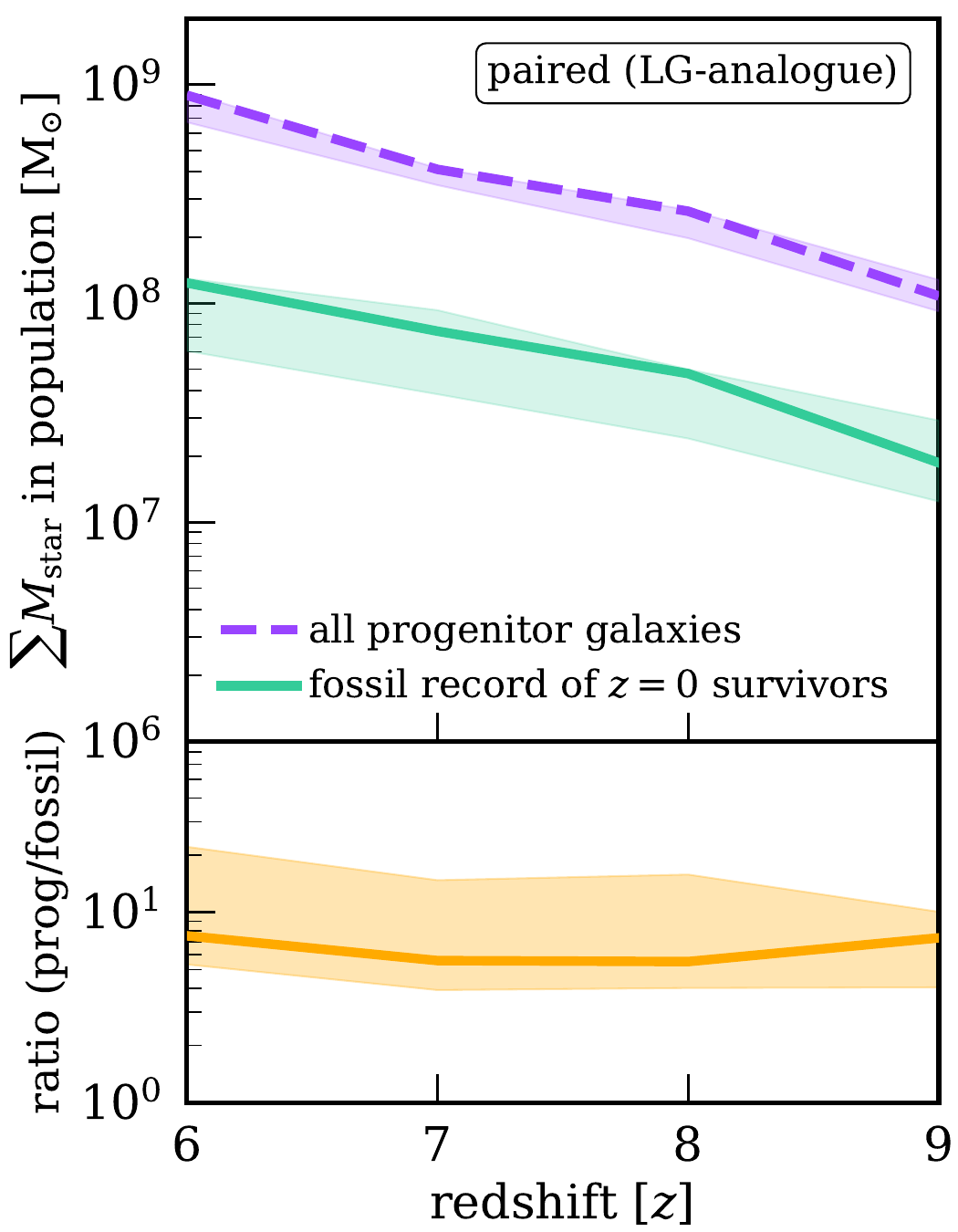} 
\end{tabular}
\caption{
\textbf{Total stellar mass in the progenitor system as a function of redshift.}
Line shows the mean across our simulations while the shaded region shows the $1\sigma$ simulation-to-simulation scatter.
\textbf{\textit{Left}}: Trends for our isolated MW-analogue simulations.
\textbf{\textit{Right}}: Same, but for the paired LG-analogue simulations.
Our selection at $z = 0$ is: all low-mass galaxies with $M_{\rm star} \geq 10^{4.5} \, \rm{M_\odot}$ out to $2$ Mpc from a MW analogue or from the geometric centre of a LG analogue.
Going back in time, we show the total stellar mass of all galaxies in the progenitor system of the present-day MW- and LG-analogue environments in purple (including galaxies that eventually form the MW/M31), and the total stellar mass from the stellar fossil record of low-mass galaxies at $z = 0$ (not including the MW/M31) in green.
The sub-panel shows the ratio of the purple to the green curves: \textit{the total mass in progenitor systems was $8 - 10 \times$ that probed by the stellar fossil record of low-mass galaxies in MW/LG environments at $z = 0$. Said differently, the total stellar mass of the progenitor systems is dominated by galaxies that eventually form the central MW/M31-mass hosts.}
}
\label{fig:mtot-populations-vs-z}
\end{figure*}

Figure~\ref{fig:mtot-populations-vs-z} compares the total cumulative stellar mass across $z = 6 - 9$ of the entire proto-MW/LG progenitor system (including galaxies that eventually form the MW/M31), compared to the total stellar mass at those redshifts probed by the fossil record of surviving low-mass galaxies at $z = 0$. Each line shows the mean across the simulations, while each shaded region shows the $1\sigma$ simulation-to-simulation scatter. Left panels show our isolated MW-analogue simulations, while right panels show the paired LG-analogue simulations; we show separate them because they have significantly different normalisations of the total mass on the y-axis. However, the overall results are similar: the total stellar mass in the entire proto-MW/LG progenitor system (including stars that end up in the MW/M31-mass galaxies at $z = 0$) is $\approx 8 - 10 \times$ the mass probed by the stellar fossil record of just the low-mass galaxies at $z = 0$. Note that in tracking back that fossil record, we consider only our sample of low-mass galaxies with $M_{\rm star} \lesssim 10^9\,\rm{M_{\odot}}$ at $z = 0$, and not the fossil record of the central MW/M31-mass galaxies.

Figure~\ref{fig:mtot-populations-vs-z} (bottom sub-panels) shows that this ratio is fairly constant across redshifts, highlighting that the stellar fossil record of low-mass galaxies that survive to $z = 0$ only probes $\approx 0.1 - 0.125 \times$ the total stellar mass of the progenitor system at $z \geq 6$. This is likely because the total stellar mass of the proto-MW/LG environment at $z \sim 6 - 9$ is dominated by the galaxies that eventually form the central MW/M31-mass hosts. We discuss the implications on the normalisation of the inferred galaxy SMF at $z \geq 6$ in Section~\ref{sec:3.x-all-z7-gal-vs-reconstruction}.

\subsection{Galaxy mass functions at $z = 7$ versus $z = 0$}
\label{sec:3.x-smf-z0-vs-z7}

\begin{figure*}
\centering
\begin{tabular}{c c}
\includegraphics[width = 0.45 \linewidth]{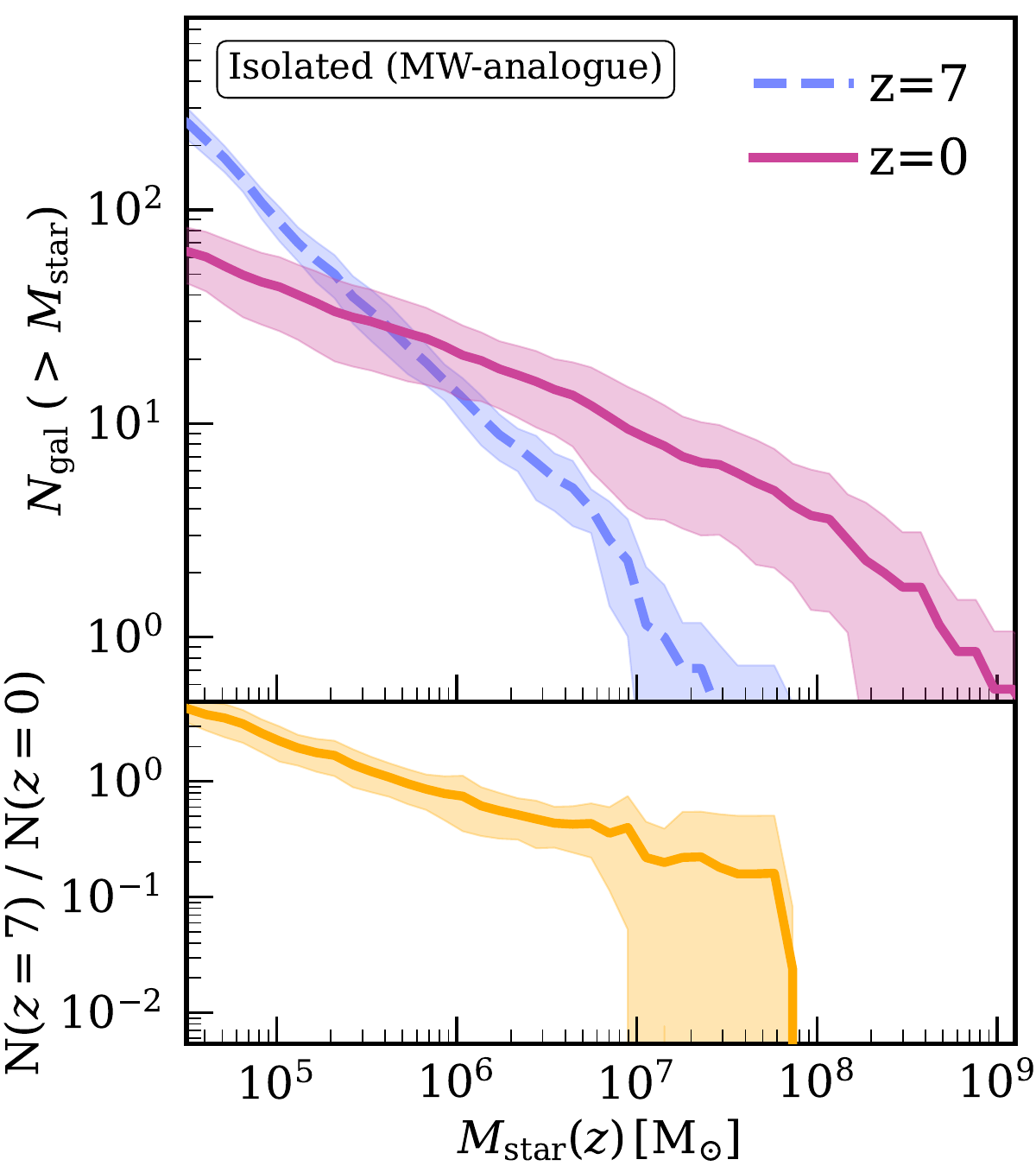}&
\includegraphics[width = 0.45 \linewidth]{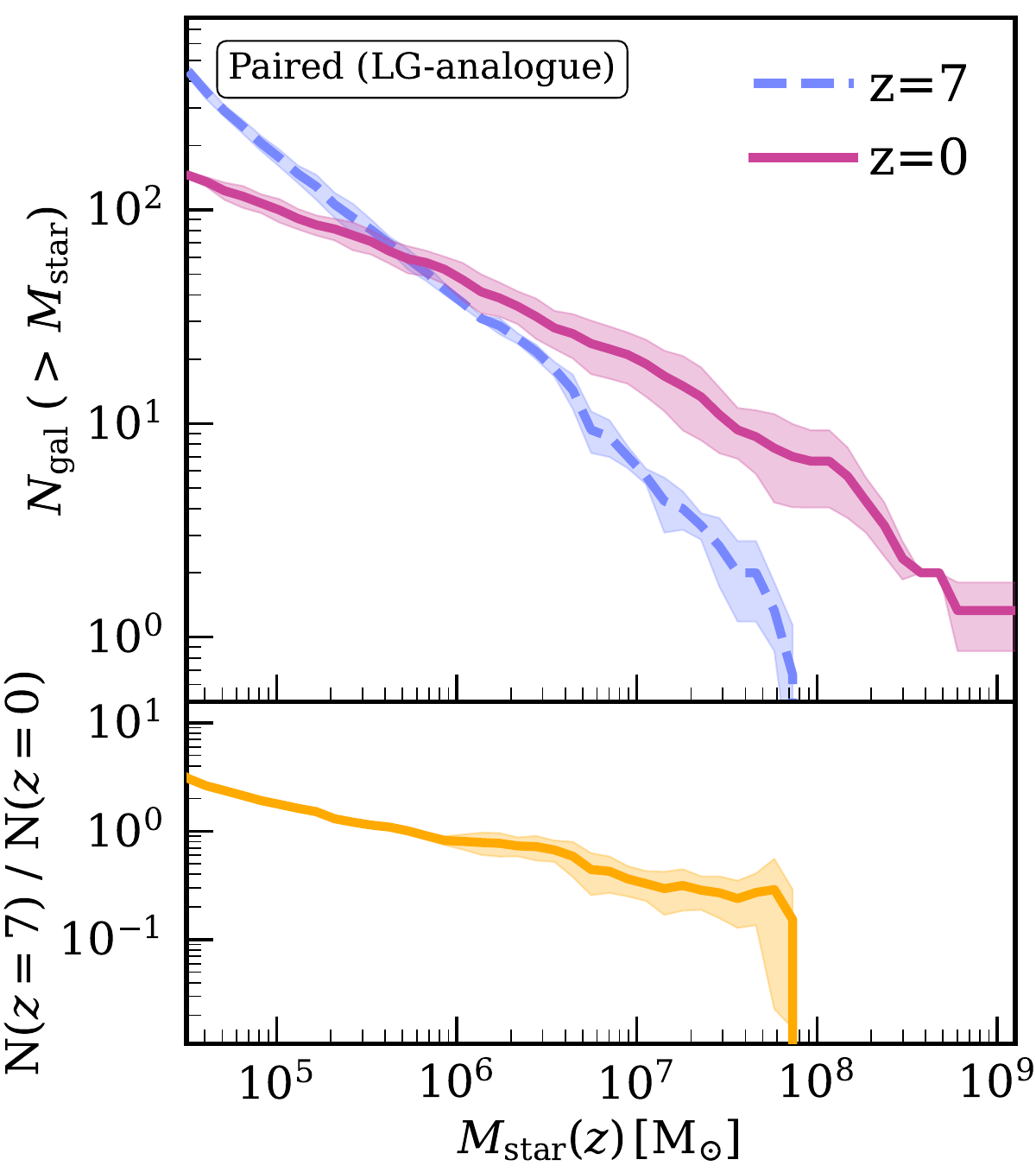}
\end{tabular}
\caption{
\textbf{Comparing cumulative galaxy stellar mass functions (SMFs) at $z = 0$ and $z = 7$}, for isolated MW-analogue simulations (left) and paired LG-analogue simulations (right).
Our selection at $z = 0$ is: all low-mass galaxies with $M_{\rm star} \geq 10^{4.5} \, \rm{M_\odot}$ out to $2$ Mpc from a MW analogue or from the geometric centre of a LG analogue (not including the MW/M31-mass central galaxy). For the SMF at $z = 7$, however, we include all progenitor galaxies, including those that eventually merged into the MW/M31-mass galaxy.
Magenta and blue curves show the mean across the simulations while the shaded regions show the $1\sigma$ simulation-to-simulation scatter. The sub-panel shows the ratio of the SMFs at $z = 7$ to $z = 0$. We compute average slopes, for the cumulative SMFs, of $\alpha (z=7) \approx -0.85$ and $\alpha (z=0) \approx -0.32$. The $z=7$ slope is much steeper because there are more low-mass galaxies and the highest-mass galaxies are less massive than at $z = 0$.
\textit{This qualitative change in the low-mass galaxy population from $z = 7$ to 0 reflects hierarchical structure formation, including galaxies growing via mergers and low-mass galaxies getting destroyed as they accrete onto the central MW/M31-mass galaxy.}
Figure~\ref{fig:smf-z0-z7-differential} shows the differential version of these SMFs.
}
\label{fig:smf-z0-z7}
\end{figure*}

To provide an initial sense of how galaxy populations in MW/LG environments vary between $z = 0$ and $z = 7$, we examine the galaxy SMFs at both redshifts. At $z = 0$ we consider all low-mass galaxies (but not the central hosts) using our fiducial selection criteria out to $2$ Mpc of a MW or LG analogue. At $z = 7$ we consider all galaxies in the progenitor systems, including those that eventually form the MW/M31. Figure~\ref{fig:smf-z0-z7} shows the cumulative SMF at $z = 0$ versus $z = 7$ for both our isolated MW analogues and paired LG analogues. Again, the trends from both suites are similar. The bottom sub-panels show the ratio of the SMFs at $z = 7$ to $z = 0$, with the orange line showing the mean. For the $1\sigma$ scatter (orange shaded region), we compute this ratio for each simulation then average across the suite.

The SMF at $z = 7$ shows a key difference from the one at $z = 0$ -- a steeper slope with a pivot point (where the two SMFs are equal) at $M_{\rm star} \sim 10^{5.5} \, \rm{M_\odot}$. By computing best-fit slopes for each simulation indvidually and then averaging across the suite, we find average values (almost identical in mean and median) of $\alpha (z=7) \approx -0.85$ and $\alpha (z=0) \approx -0.32$ for the cumulative SMFs, and $\alpha (z=7) \approx -1.85$ and $\alpha (z=0) \approx -1.32$ for the differential ones. This is a natural consequence of hierarchical structure formation of galaxies: low-mass galaxies in the proto-LG merge with each other to grow and can also accrete onto the central MW/M31-mass galaxy. \citet{santistevan-20} showed, also using the FIRE-2 simulations, that the surviving low-mass galaxy population around MW-mass galaxies at $z = 0$ is a highly incomplete census of the low-mass galaxies that existed in the progenitor system that built up the MW/LG. They also showed that the redshift at which the number of galaxies at a given stellar mass peaks is lower for higher mass, which is another natural consequence of hierarchical structure formation.
\textit{This hierarchical assembly, including mergers and disruptions, is precisely what might affect the inference of the high-redshift SMF/UVLF using the near-far reconstruction method (because tracking star formation histories contains no information about mergers), and we characterise this effect in the rest of this paper.}

Appendix~\ref{appA} and Figure~\ref{fig:smf-z0-z7-differential} show the differential (instead of cumulative) versions of these SMFs, with the same overall takeaways.

\subsection{Progenitor galaxies at $z = 7$}
\label{sec:3.x-prog-at-z7}

\subsubsection{How many high-redshift progenitors do galaxies at $z=0$ have?}
\label{sec:3.x.x-how-many-z7-progs}

\begin{figure*}
\centering
\begin{tabular}{c c}
\includegraphics[width = 0.45 \linewidth]{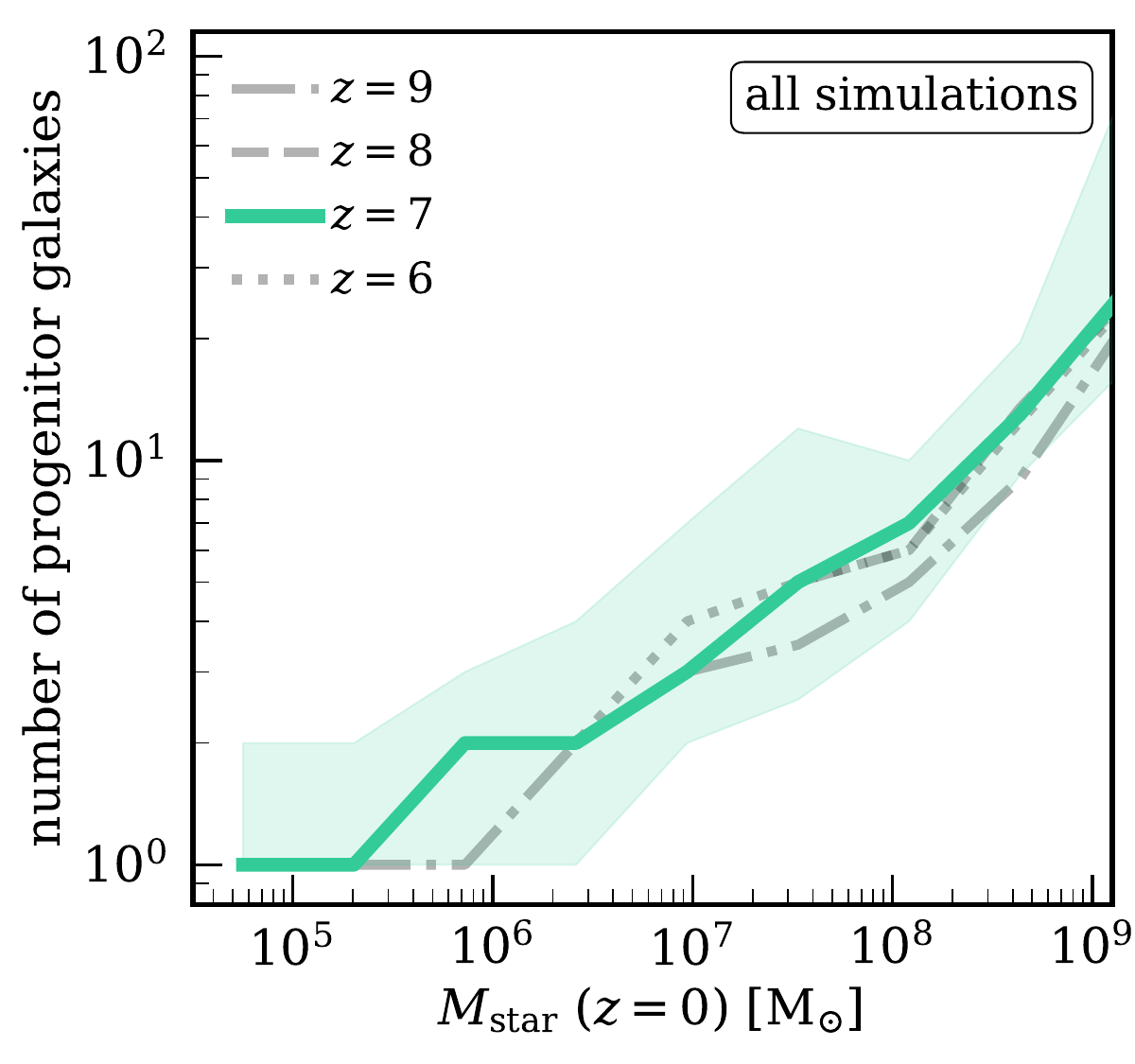} &
\includegraphics[width = 0.45 \linewidth]{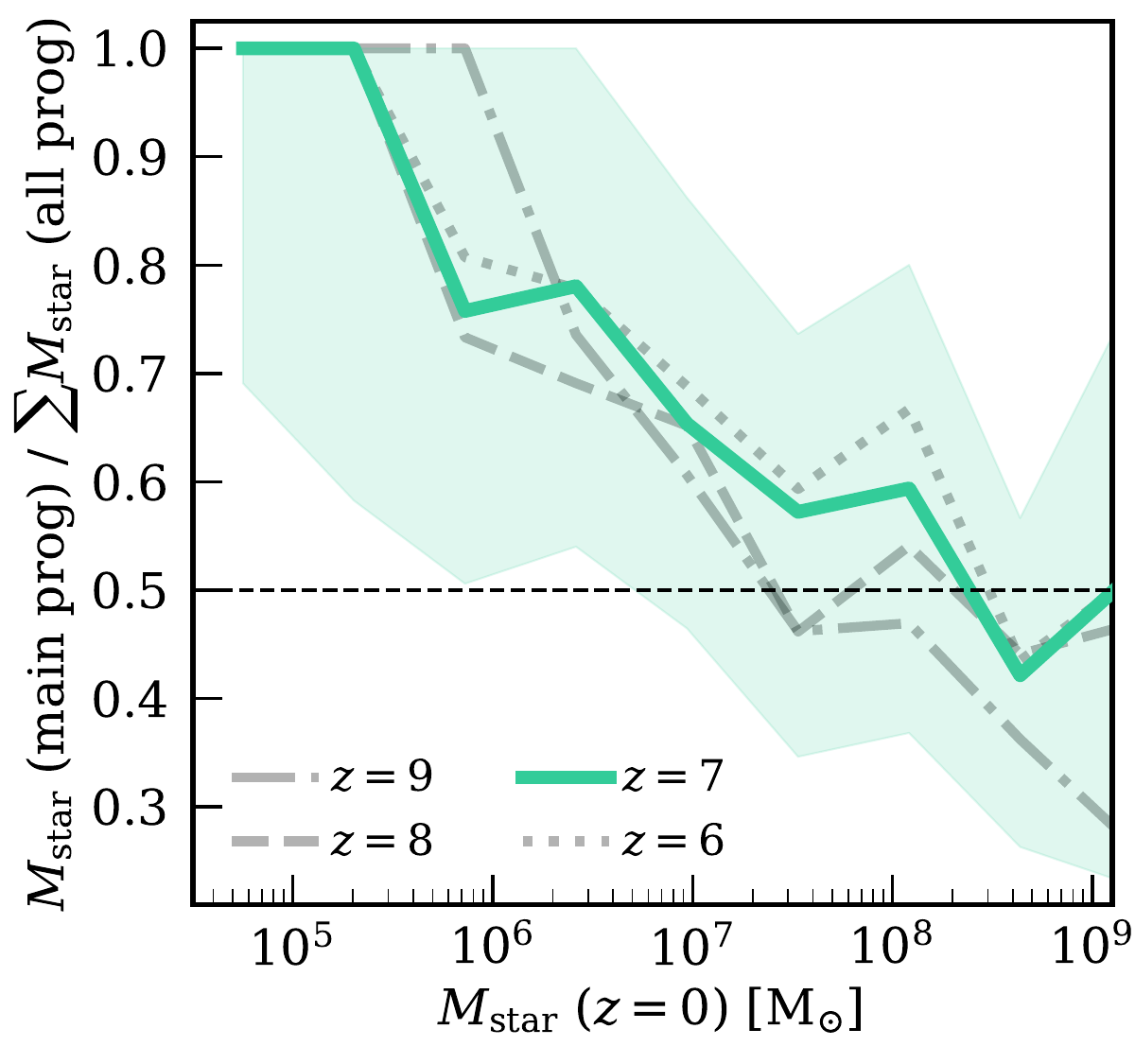}
\end{tabular} 
\caption{
\textbf{Total number of galaxy progenitors (with $M_{\rm star} > 10^{4} \, \rm{M_\odot}$) at $z > 6$ (left) and the stellar mass contribution from the main stellar progenitor (right) for all low-mass galaxies within 2 Mpc of a MW or LG analog at $z = 0$.} We find no significant difference between MW- and LG-analogue environments, so we combine them here.
The solid green curve shows the median across all simulations at $z = 7$, and the shaded region shows the $68$ per cent simulation-to-simulation scatter. We also find little evolution across $z = 6 - 9$ in both the median and scatter (latter not shown).
\textbf{\textit{Left}}: The number of high-redshift progenitor galaxies increases with present-day stellar mass. Ultra-faint galaxies had only $1 - 2$ progenitor galaxies (with $M_{\rm star} > 10^{4} \, \rm{M_\odot}$) at $z \gtrsim 6$, while LMC-mass galaxies with $M_{\rm star}(z = 0) \sim 10^9 \, \rm{M_\odot}$ had $\sim 15 - 40$ progenitors.
\textbf{\textit{Right}}: The fractional contribution of the main progenitor to the overall stellar mass (from all progenitors) at a given redshift of a galaxy with a given $M_{\rm star}(z = 0)$. This fraction decreases with present-day stellar mass: for ultra-faint galaxies today, the main progenitor contributed $\approx 100\%$ to the total progenitor stellar mass at $z \gtrsim 6$, and for LMC-mass galaxies today, the main progenitor contributed $\approx 50\%$.
\textit{Thus, near-far reconstruction applied to galaxies with $M_{\rm star}(z = 0) \lesssim 10^6 \, \rm{M_\odot}$ is particularly straightforward, given that they typically had $1 - 2$ progenitors.
Furthermore, while more massive galaxies had a larger number of progenitors, at all masses up to LMC mass, the typical main progenitor almost always contributes the majority of progenitor stars.}
}
\label{fig:num-z7-allprog-for-z0-gals}
\end{figure*}

To address the question of mergers and disruption of progenitor low-mass galaxies in the proto-MW/LG, we investigate how many progenitors the present-day galaxies had at different redshifts. Figure~\ref{fig:num-z7-allprog-for-z0-gals} (left panel) shows the median number of $z = 7$ progenitors that a low-mass galaxy at $z = 0$ in our entire suite of simulations has as the solid green curve, with the light green shaded region showing the $16 - 84$th percentile scatter. We also show the median for $z = 6$, $8$, and $9$; there is qualitatively no change with redshift. For legibility, we do not show shaded regions for any redshift besides $z = 7$, because the scatter does not qualitatively change with redshift.

As Figure~\ref{fig:num-z7-allprog-for-z0-gals} (left panel) shows, present-day ultra-faint galaxies with $M_{\rm star} \sim 10^5 \, \rm{M_\odot}$ have at most $\approx 1 - 2$ high-redshift progenitors \citep[see also][]{fitts-18}, while galaxies like the Large Magellanic Cloud (LMC) with $M_{\rm star} \sim 10^9 \, \rm{M_\odot}$ have $\approx 15 - 40$ high-redshift progenitors. Although this rise in the average number of progenitors with the present-day stellar mass of galaxies is also a natural consequence of hierarchical structure assembly, the number of progenitors at $z = 7$ rises quickly with increasing mass at $z = 0$, which SFH reconstruction does not reflect, because (on its own) it provides no information about past mergers or the number of progenitors.

\subsubsection{Contribution of the main progenitor galaxy}
\label{sec:3.x.x-how-many-mp-z7}

SFH reconstruction, and therefore the near-far technique, inherently assumes that one progenitor contained all of the stellar mass at any given redshift. Since we demonstrate that galaxies at $z = 0$ can have numerous progenitors at $z \geq 6$, we investigate the importance of the single main stellar progenitor. Figure~\ref{fig:num-z7-allprog-for-z0-gals} (right panel) shows, for our sample of low-mass galaxies at $z = 0$, the stellar mass of the single main progenitor relative to the total stellar mass in all progenitor galaxies at that redshift. The solid green curve shows the median at $z = 7$, with the $16 - 84$th percentile scatter as the green shaded region. The light grey curves show the median at the other redshifts. While the median trend does not show significant evolution with redshift, there is a slight downward shift with increasing redshift for galaxies with $M_{\rm star} \gtrsim 10^7 \, \rm{M_{\odot}}$. This is likely because the main progenitor is slightly less dominant at higher redshifts, and with more mergers in the intervening period. As with the left panel, we do not show shaded regions for the other redshifts for legibility, because there is no significant variation with redshift.

The mass fraction of the main progenitor at $z = 7$ relative to the total mass in all progenitors is at almost $100$ per cent for the lowest-mass galaxies at $z = 0$. This makes sense, because these galaxies only had $\approx 1 - 2$ progenitors at $z \sim 6 - 9$. For galaxies at the high-mass end of our sample, the main progenitor contributed on average $50$ per cent of the total progenitor stellar mass at $z = 7$. Although this is lower than for our lowest-mass galaxies, it still generally dominates the overall progenitor mass budget, implying that across our entire mass range at $z = 0$, the single main stellar progenitor dominates the total progenitor mass budget.
\textit{This result offers confidence in the ability of SFH reconstruction and the near-far technique to infer the \textbf{slope} of the SMF at high-redshift, because the assumption/approximation that a single main progenitor dominated the mass budget at $z \geq 6$ is reasonable for galaxies with $M_{\rm star}(z = 0) \sim 10^{4.5} - 10^9 \, \rm{M_\odot}$, at least up to redshifts of $z \sim 8 - 9$.} This result also agrees with that from \citep{fitts-18}, who showed similar trends for FIRE galaxies with present-day masses of $10^5 \lesssim M_{\rm star} \lesssim 10^7 \, \rm{M_{\odot}}$ .

\subsection{How well the stellar fossil record today recovers the SMF of progenitor galaxies}
\label{sec:3.x-all-z7-gal-vs-reconstruction}

Here we present the key analysis and results of this paper: \textit{testing directly how well the near-far technique infers the slope of the SMF at the low-mass end at $z \sim 6 - 9$, assuming perfect CMD-based reconstruction of the SFHs of low-mass galaxies at $z = 0$.} To test how well the stellar fossil record of low-mass galaxies at $z = 0$ recovers the overall SMF of the progenitor system at $z = 7$, we compare the following galaxy populations: (a) all galaxies in the progenitor system including those that eventually form (end up in) the MW/M31-mass galaxy, (b) all progenitor galaxies of surviving low-mass galaxies at $z = 0$ (except the central MW/M31-mass host), (c) main progenitor galaxies of surviving low-mass galaxies at $z = 0$, and (d) the populations probed by the stellar fossil record of surviving low-mass galaxies at $z = 0$.

\begin{figure*}
\centering
\begin{tabular}{c c}
\includegraphics[width = 0.48 \linewidth]{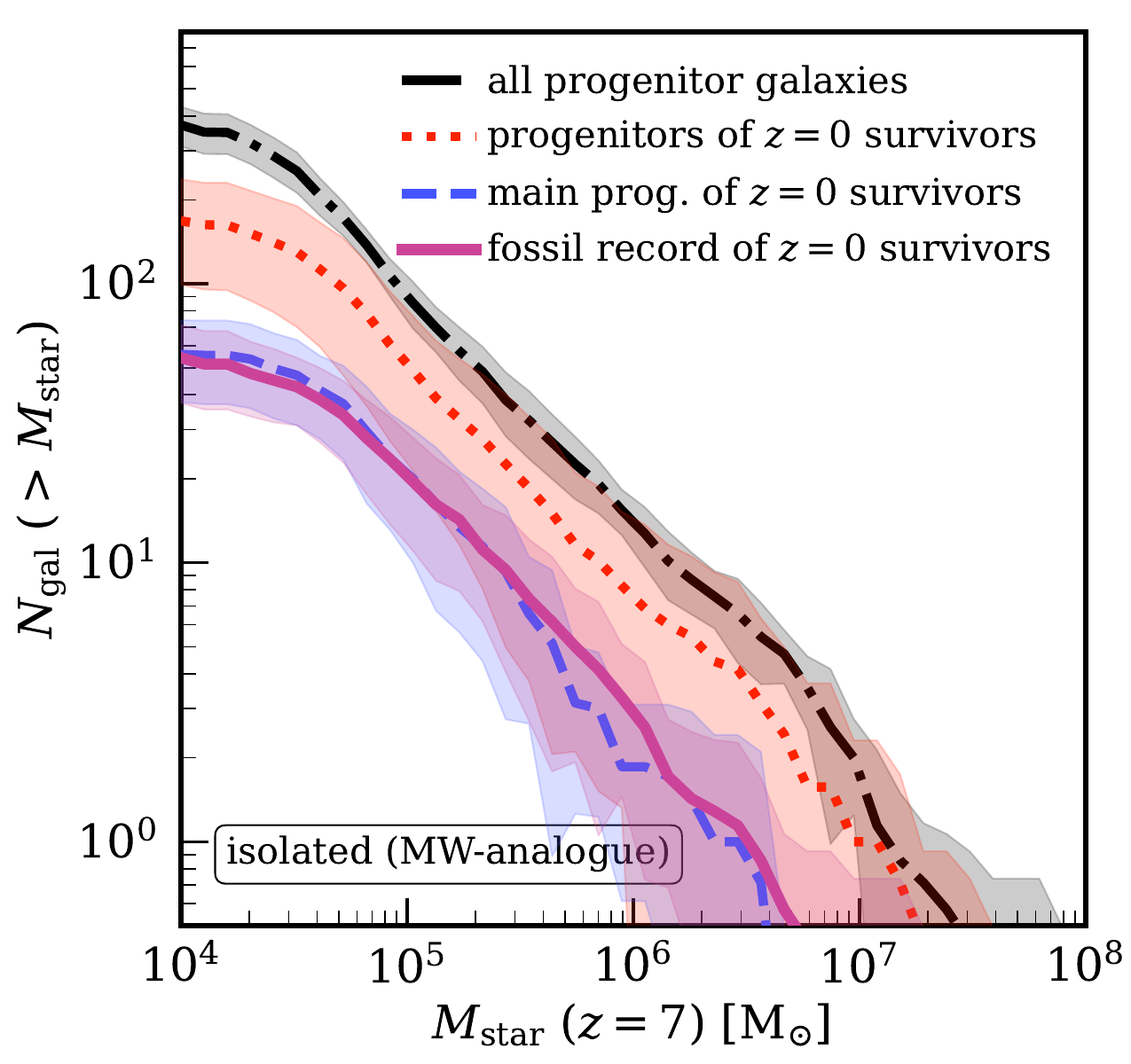}&
\includegraphics[width = 0.48 \linewidth]{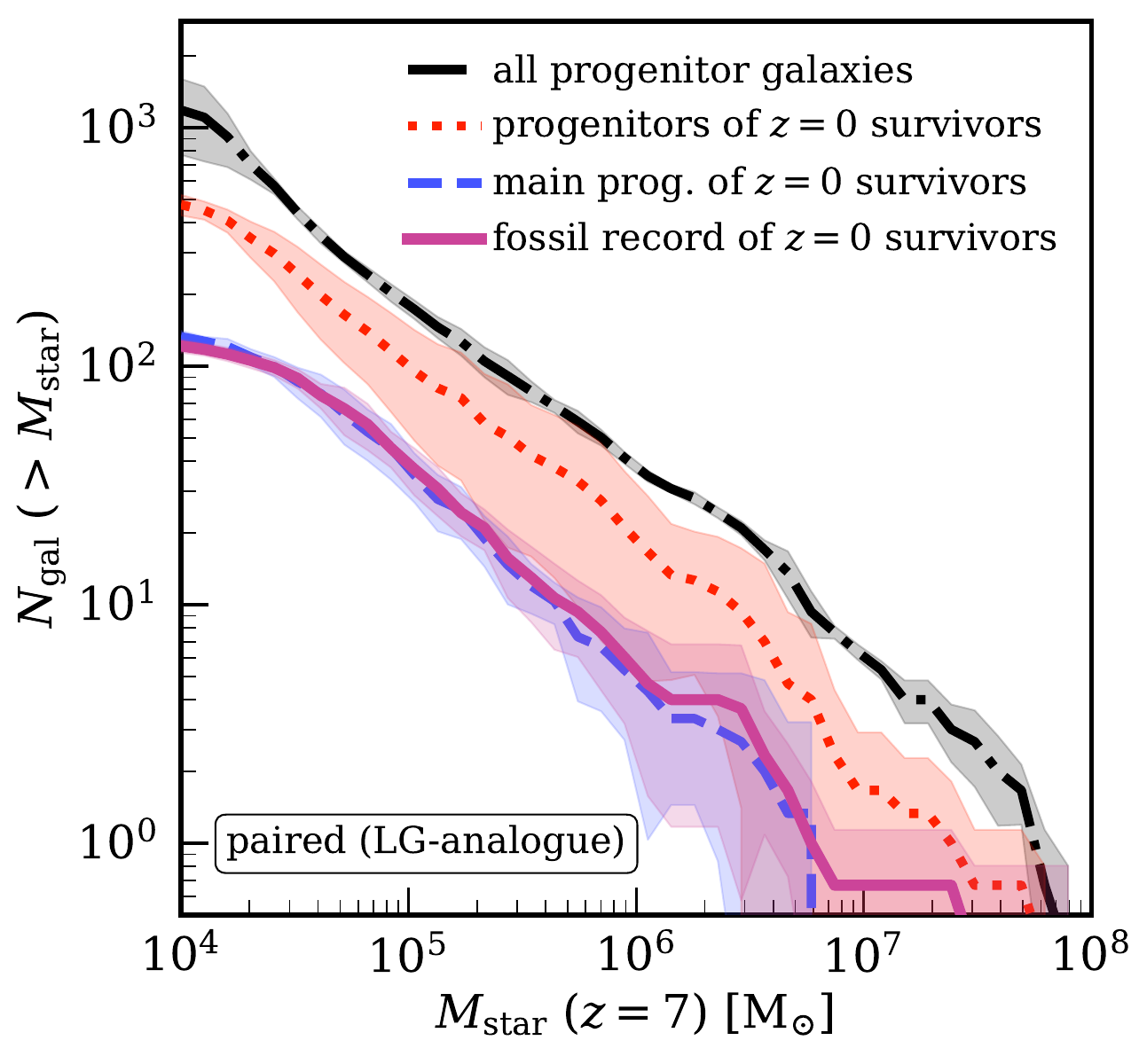}\\
\end{tabular}
\includegraphics[width = 0.75 \linewidth]{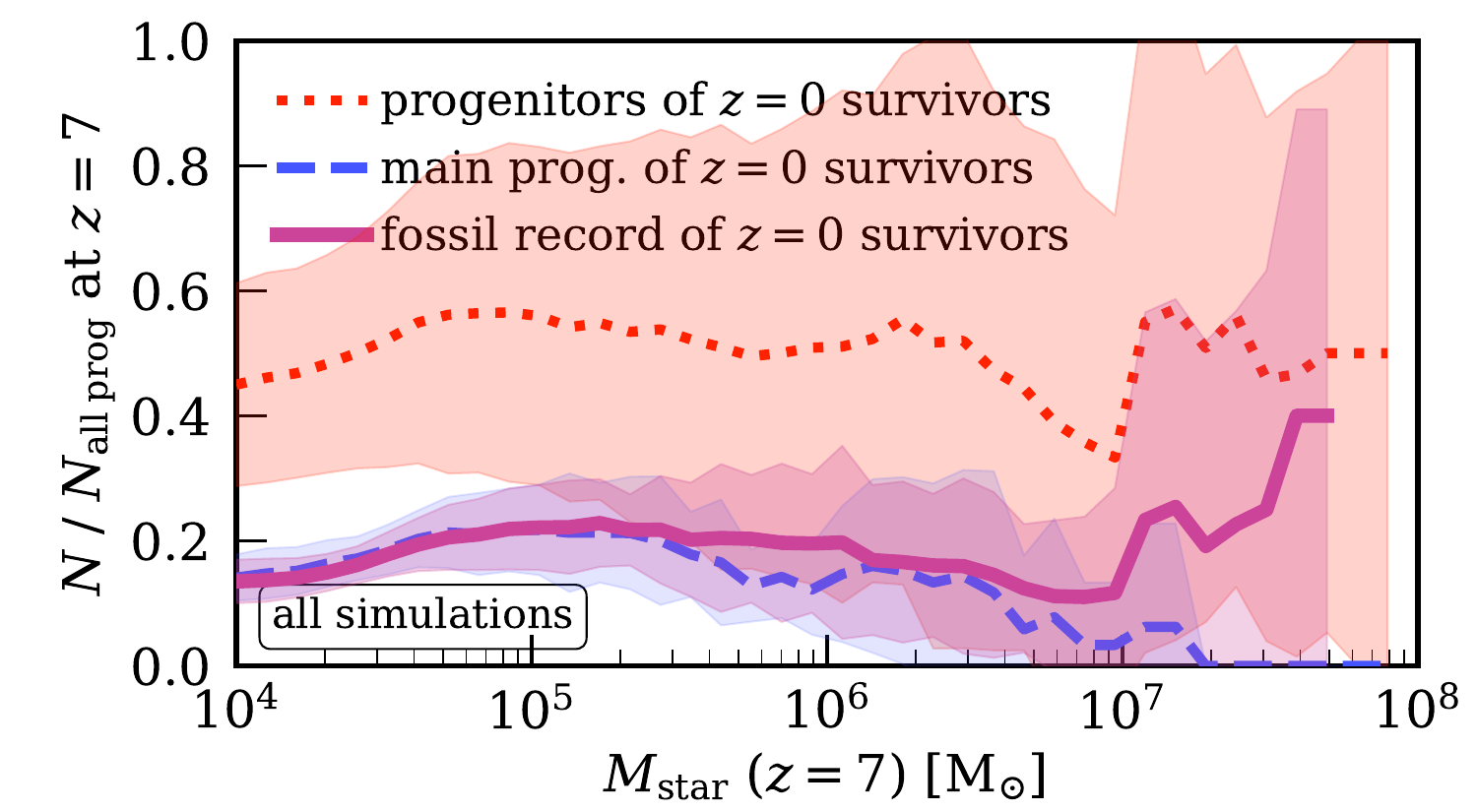}
\caption{
\textbf{The accuracy of near-far reconstruction of the stellar mass function (SMF) at $z = 7$}, using the fossil record of all low-mass galaxies at $z = 0$ out to $2$ Mpc from a MW- or LG-analogue, excluding the MW/M31-mass host galaxy.
\textbf{\textit{Top row}}: The cumulative SMF at $z = 7$ for the following galaxy populations: (a) every galaxy in the progenitor system, including those that eventually form (end up in) the MW/M31-mass galaxy (dashdot black), (b) all progenitors of surviving low-mass galaxies at $z = 0$ (dotted red), (c) main progenitors of surviving galaxies at $z = 0$ (dashed blue), and (d) the fossil record of surviving galaxies at $z = 0$, assuming that all stellar mass at $z = 7$ was in a single progenitor, as one would infer from the SFH at $z = 0$ (solid magenta).
Curves show the mean across simulations and shaded regions show the $1\sigma$ simulation-to-simulation scatter, for the isolated MW-analogue (left) and LG-analogue (right) environments.
\textbf{\textit{Bottom}}: Ratio of each progenitor sub-population (colored curves) to the total progenitor population (black curve).
In this panel, we show results for all the simulations combined, because there is no significant difference in normalisation between the MW-analogues and LG-analogues.
\textit{The magenta curve shows the `near-far' approach of using the SFHs from the stellar fossil record at $z = 0$: while the total number of galaxies inferred at $z = 7$ is only $15 - 20\%$ that of the true number of all progenitor galaxies, the key result is that this ratio is nearly flat, which means that the `near-far' approach can recover the low-mass end slope/shape in an unbiased manner.}
Figure~\ref{fig:different-smfs-compared-to-all-at-z7-differential} shows the differential version of these SMFs and ratios, with the same takeaways.
}
\label{fig:different-smfs-compared-to-all-at-z7}
\end{figure*}

Figure~\ref{fig:different-smfs-compared-to-all-at-z7} (upper panels) shows the cumulative SMFs for all four populations. We show this separately for the isolated MW-analogue environments and the paired LG-analogue environments, because the normalisation of the total number of galaxies is quite different in the two cases, leading to different y-axis dynamic ranges. Each curve shows the mean across the simulations, while each shaded region shows the $1\sigma$ simulation-to-simulation scatter. The overall shape of all four SMFs is qualitatively the same. The apparent flattening at the lowest masses ($\approx 10^4 \, \rm{M_\odot}$) arises from the mass resolution limit of our simulations, and is not proof of a physical rollover or flattening in the SMFs.

To qualitatively compare the shapes/slopes of the SMFs, Figure~\ref{fig:different-smfs-compared-to-all-at-z7} (bottom panel) shows the ratios of the red, blue, and magenta curves to the black curve. We compute ratios for each simulation first and then average them across all simulations. The trends are similar for both the isolated MW-analogue and paired LG-analogue simulations, so we combine them in a single panel. All three curves in the bottom panel are fairly flat with stellar mass, although the scatter increases towards higher masses due to the smaller numbers of higher-mass galaxies in our sample. The ratio of the fossil record SMF to the true SMF (magenta curve in the bottom panel) is fairly constant from $M_{\rm star} \sim 10^4 - 10^{7.5} \, \rm{M_\odot}$, which implies that the low-mass end slope of the SMF inferred using the stellar fossil record is similar to the slope of the `true' SMF of all galaxies in proto-MW/LG-like environments at $z = 7$. This key result provides confidence in the ability of the near-far approach to recover the slope of the low-mass end of the SMF at $z \approx 7$. To provide a more quantitative demonstration of this agreement, we also discuss a comparison of the best-fit slopes later on in this section and in Figure~\ref{fig:slope-comparison}.

To first order, the slope/shape of the low-mass end of the SMF at high-redshift is most important when considering the question of what reionisation-era galaxy populations looked like, and whether low-mass galaxies were dominant in driving reionisation compared to more massive ones. However, a secondary consideration is that of the normalisation (total number of galaxies) of the high-redshift SMF, and Figure~\ref{fig:different-smfs-compared-to-all-at-z7} (bottom panel) shows that the stellar fossil record of galaxies at $z = 0$ only recovers $15 - 20$ per cent of the `true' total number of galaxies in the proto-MW/LG progenitor system at $z = 7$. This is a natural consequence of the mergers and disruption of low-mass galaxies, wherein they can merge amongst themselves and can also accrete onto the central MW/M31-like host galaxy. The dotted red line for all progenitors of survivors at $z = 0$, in the bottom panel suggests that $\approx 50$ per cent of the number of low-mass galaxies in the progenitor system at $z = 7$ merge into the MW/M31-mass galaxy by $z = 0$, and are therefore not recoverable using the fossil record of low-mass galaxies surviving at present day. \textit{However, because this accretion is fairly mass-independent as the relatively flat nature of the dotted red curve shows, it does not bias the slope of the SMF at $z = 7$ inferred using the stellar fossil record of surviving low-mass galaxies at $z = 0$.} Finally, this bottom panel of Figure~\ref{fig:different-smfs-compared-to-all-at-z7} highlights the following: $\approx 50$ per cent of low-mass galaxies in the proto-LG environment at $z=7$ merge into the central host, $\approx 15-20$ per cent of them show up in the fossil record of surviving low-mass galaxies at $z=0$, and the remaining $30-35$ per cent account for the numerous progenitors of surviving low-mass galaxies at $z=0$, whose numbers are not traced by the fossil record.

Appendix~\ref{appB} shows the differential (instead of cumulative) version of these SMFs, with the same overall takeaways.
Furthermore, Appendix~\ref{appC} shows the result for the fossil record curve in Figure~\ref{fig:different-smfs-compared-to-all-at-z7} (bottom), but for each simulation separately, instead of averaging across the entire suite. This demonstrates that this result is relatively robust for individual simulations as well, and is not simply a consequence of averaging across our suite.

\begin{figure*}
\centering
\begin{tabular}{c}
\includegraphics[width = 0.75 \linewidth]{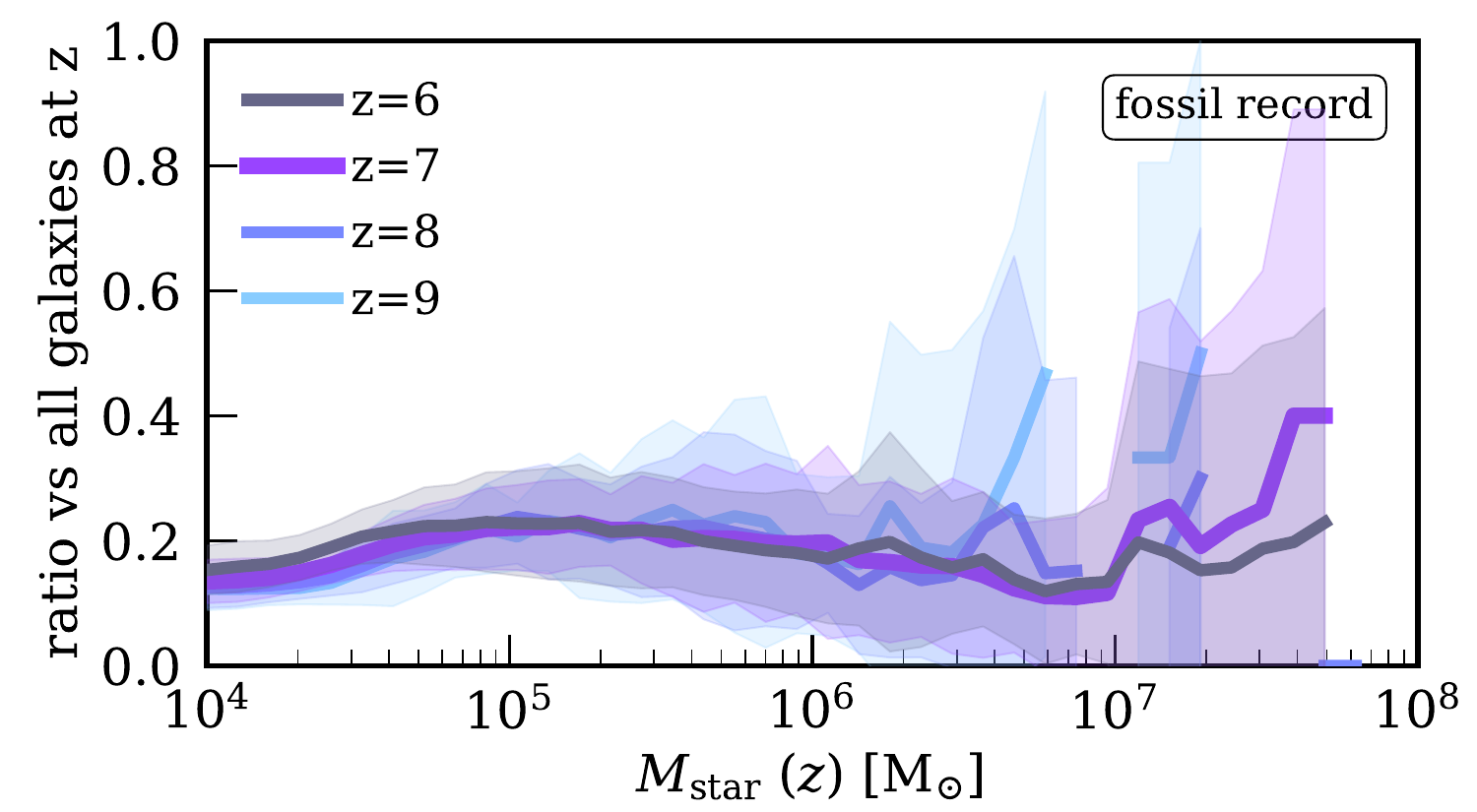} 
\end{tabular} 
\caption{
\textbf{The same as the solid magenta curve in Figure~\ref{fig:different-smfs-compared-to-all-at-z7} (bottom), but showing redshifts from $z = 6$ to $z = 9$}, in all cases using the SFH from the stellar fossil record of surviving galaxies at $z = 0$ to infer $M_{\rm star}(z)$, assuming all stellar mass was in a single progenitor. \textit{We find qualitatively no change from $z = 6 - 9$, so the `near-far' approach is robust and unbiased for recovering the slope of the SMF at the low-mass end across the likely redshift range of reionization.}
}
\label{fig:redshift-dependence}
\end{figure*}

We perform the same tests at $z = 6$, $8$, and $9$, to study the validity of the near-far technique across a larger time period during the Epoch of Reionisation. Figure~\ref{fig:redshift-dependence} shows how the magenta curve (ratio of the fossil record inference to the overall progenitor population) in the bottom panel of Figure~\ref{fig:different-smfs-compared-to-all-at-z7} behaves as a function of progenitor redshift. The mean and scatter are qualitatively the same across the entire redshift range. This suggests that SFH reconstruction and the stellar fossil record are able to recover the low-mass end slope of the SMF at $z = 6 - 9$ fairly accurately, while recovering $15 - 20$ per cent of the total number of galaxies (that is, the normalisation of the SMF). The result at $z = 9$ (lightest blue curve) is the least robust, likely in part because of the smaller sample of galaxies in our simulations at such early time.

\begin{figure}
\centering
\begin{tabular}{c}
\includegraphics[width = 0.98 \linewidth]{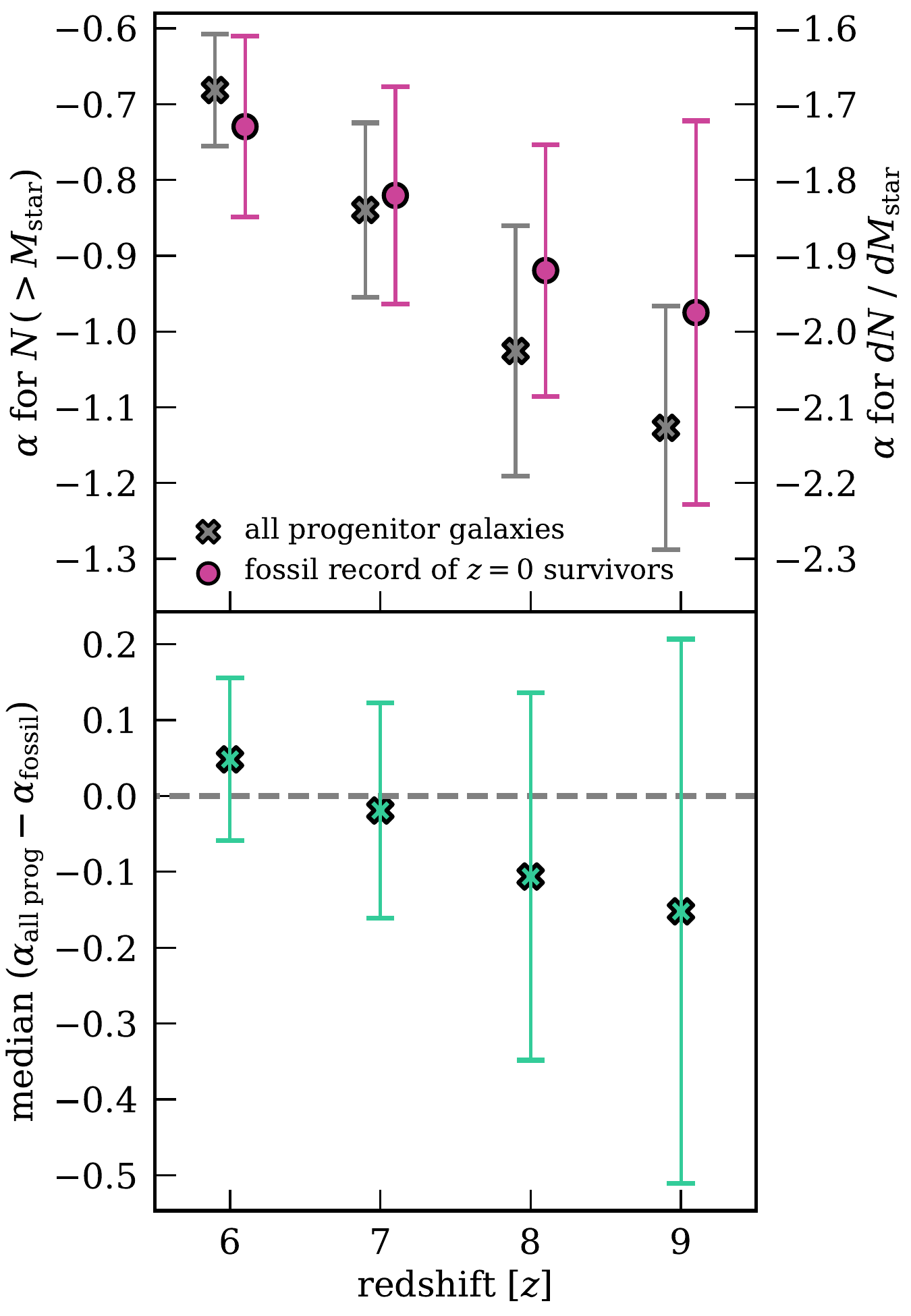} 
\end{tabular} 
\caption{
\textbf{The accuracy of history-based reconstruction of the low-mass end slopes of the SMF, at different redshifts.} \textbf{\textit{Top:}} best-fit slopes for the SMF of all galaxies in the progenitor system (dashdot black curves from Figure~\ref{fig:different-smfs-compared-to-all-at-z7} top panels) as black crosses, versus those for the SMF derived from the fossil record (solid magenta curves from Figure~\ref{fig:different-smfs-compared-to-all-at-z7} top panels) as magenta circles, across $z = 6 - 9$. Each point shows the median value across all simulations, and the error bar shows the $68$ per cent simulation-to-simulation scatter. The left y-axis shows the slope for the cumulative SMF while the right y-axis shows the slope for the corresponding differential SMF. Finally, the grey dashed line at a difference of zero is to guide the eye. \textbf{\textit{Bottom:}} a more direct comparison, showing the median and $68$ per cent scatter of the \textit{difference} in slope between the SMF of all galaxies and the SMF inferred from the fossil record. The difference is computed first for each simulation separately and then averaged. \textit{Both panels show good agreement within $1\sigma$ between the two slopes, especially at $z = 6$ and $z = 7$. Although the agreement is worse at higher redshifts, it is still within $1\sigma$ simulation-by-simulation scatter, and the median difference in slope remains within $\approx 0.15$ up to $z = 9$.}
}
\label{fig:slope-comparison}
\end{figure}

As a more quantitative comparison of slopes, Figure~\ref{fig:slope-comparison} compares of best-fit low-mass slopes from (a) the SMFs of all galaxies in the progenitor system (black crosses; same as dashdot black curve in Figure~\ref{fig:different-smfs-compared-to-all-at-z7}), and (b) the SMFs inferred from the fossil record of surviving low-mass galaxies at $z = 0$ (magenta circles; same as the solid magenta curve in Figure~\ref{fig:different-smfs-compared-to-all-at-z7}). We measure the low-mass slope by fitting a power law to the SMF for each simulation across $M_{\rm star} = 10^{4.5} - 10^{6.5} \, \rm{M}_\odot$. We choose this mass range to avoid the unphysical flattening of the SMFs at $M_{\rm star} \lesssim 10^{4.5} \, \rm{M_\odot}$ from the resolution limit of our simulations, and to avoid the turnover at higher masses where the SMF transitions from power-law to exponential behaviour, assuming a Schechter function shape.

Figure~\ref{fig:slope-comparison} (top) shows the best-fit slopes for both populations versus redshift, averaged across all simulations. We show slopes fit to the cumulative SMF on the left y-axis and those independently fit to the differential SMF on the right y-axis. We find that although the agreement between the slope of the `true' SMF and the SMF inferred from the fossil record (as an average across our simulations) is best at $z = 6 - 7$ and gets somewhat worse with increasing redshift, the median slopes still agree within $68$ per cent scatter, adding further confidence to the validity of near-far technique. Additionally, we find that the SMFs steepen with increasing redshifts, with the average slopes for the cumulative SMFs going from: (a) $\alpha \approx -0.68$ at $z \approx 6$ to $\alpha \approx -1.13$ at $z \approx 9$ for the SMF of all galaxies in the progenitor system, and (b) $\alpha \approx -0.73$ at $z \approx 6$ to $\alpha \approx -0.98$ at $z \approx 9$ for the SMF inferred using fossil record reconstruction. The average slopes of the corresponding differential SMFs go from: (a) $\alpha \approx -1.68$ at $z \approx 6$ to $\alpha \approx -2.13$ at $z \approx 9$ for the SMF of all progenitor galaxies, and (b) $\alpha \approx -1.73$ at $z \approx 6$ to $\alpha \approx -1.98$ at $z \approx 9$ for the fossil-inferred SMF. We verify by independently computing differential SMFs and their average slopes, that the relation $\alpha_{\rm diff} = \alpha_{\rm cumul} - 1$ still holds true. 

Since the objective of this study is to stress-test the accuracy of the near-far technique, rather than commenting on the exact low-mass slope of the SMF at $z \gtrsim 6$, Figure~\ref{fig:slope-comparison} (bottom) shows a more direct and self-consistent comparison, via the difference in slopes between the SMFs of the two populations at a given redshift. We compute the difference individually within each simulation first and then compute median values and $68$ per cent scatter across all simulations. At $z \sim 6 - 7$, the median difference in slopes is $\approx 0$, with the scatter being within $\approx 15$ per cent. \textit{Although the median difference increases with increasing redshift, it is still only up to $\approx 0.15$ even at $z = 9$, which demonstrates the accuracy of near-far reconstruction.}

\section{Summary, Caveats \& Discussion}
\label{sec:4-conclusions}

\subsection{Summary of key results}
\label{sec:4.x-key-results}

\begin{enumerate}
\item \textit{Primary significance of the present-day stellar fossil record:} the fossil record of surviving low-mass galaxies in the LG at $z = 0$ probes $\approx 0.1 - 0.125 \times$ the total stellar mass of all galaxies in the progenitor system (including the progenitors of the MW/M31) at $z = 6 - 9$ (Figure~\ref{fig:mtot-populations-vs-z}). Said differently, $\approx 90$ per cent of the stellar mass (and $\approx 50$ per cent of the number of galaxies) in the proto-MW/LG environment ends up forming (gets accreted into) the central MW/M31-like host galaxy.
\item \textit{Comparing galaxy stellar mass functions (SMF) at $z = 0$ and $z = 7$:} the SMF at $z = 7$ is much steeper ($\alpha_{\rm cumul} \approx -0.85$) than at $z=0$ ($\alpha_{\rm cumul} \approx -0.32$). This reflects the hierarchical formation and growth of galaxies, with lower-mass systems merging to form higher-mass ones over cosmic time (Figures~\ref{fig:smf-z0-z7}  and \ref{fig:smf-z0-z7-differential}).
\item \textit{Number of high-redshift progenitors and importance of the main progenitor:} for galaxies at $z = 0$, the average number of progenitor galaxies at $z = 7$ rises sharply with present-day stellar mass, going from $\approx 1 - 2$ for ultra-faint galaxies to $\approx 30$ for LMC-mass galaxies. The fractional stellar mass contribution of the main progenitor to the overall mass of all progenitors at $z = 7$ drops from $\approx 100$ per cent to $\approx 50$ per cent from the ultra-faint to the LMC-mass regimes, but the main progenitor still usually dominates the high-redshift population at all masses we consider. These results are similar for galaxies in MW- and LG-analogue environments, and fairly consistent from $z = 6 - 9$ (Figure~\ref{fig:num-z7-allprog-for-z0-gals}).
\item \textit{Near-far reconstruction accurately infers the slope/shape of the SMF at low masses, without significant bias, at $z \sim 6 - 9$.} Assuming perfect SFH reconstruction using CMD modelling and using the stellar fossil record of low-mass galaxies at $z = 0$ in MW- and LG-analogue environments, we show recovery of $15 - 20$ per cent of the total number of all progenitor galaxies at $z \geq 6$. More importantly however, the inferred slope/shape at $z \sim 6 - 9$ is accurate (within $1 \sigma$) at $M_{\rm star} \lesssim 10^{6.5} \, \rm{M_\odot}$, at least down to $M_{\rm star} \approx 10^{4.5} \, \rm{M_\odot}$, the resolution limit of our simulations (Figures~\ref{fig:different-smfs-compared-to-all-at-z7}, \ref{fig:redshift-dependence}, \ref{fig:slope-comparison}, \ref{fig:different-smfs-compared-to-all-at-z7-differential}, and \ref{fig:different-smfs-compared-to-all-at-z7-individual}). 
\end{enumerate}

\subsection{Caveats}
\label{sec:4-caveats}

\subsubsection{How representative is the proto-Local Group of the overall universe at $z > 6$?}
\label{sec:4.x-lg-representativeness}

We show that the stellar fossil record of low-mass galaxies in the LG at $z = 0$ provides an unbiased inference of the slope of the SMF of the LG's progenitor system at $z \geq 6$. However, the concern remains: does the progenitor system of the LG accurately represent the overall galaxy population at high redshift? This raises the question of whether the volume of the MW/LG's progenitor system at $z \sim 6 - 9$ is large enough to be typical of the global galaxy population at those epochs. \citet{boylan-kolchin-16} show, using the \textsc{ELVIS} suite of dark matter-only cosmological zoom-in simulations \citep{garrison-kimmel-14}, that the proto-LG spanned a comoving volume of $\sim 350$ Mpc$^3$ at $z \approx 7$; equivalent to the volume of the Hubble Ultra-Deep Field (HUDF) at those redshifts. They also showed that the DM halo mass function (HMF) of the proto-LG at $z \approx 7$ agrees well with the cosmological expectation from the Sheth-Tormen mass function \citep{sheth-01}. 

We will further dig into this question in a future Gandhi et al., paper using our baryonic FIRE simulations, by comparing the SMFs at $z \sim 6 - 9$ that we infer from our MW- and LG-analogue simulations to the SMFs from the high-redshift suite of FIRE-2 simulations \citep[introduced in][]{ma-18-a, ma-19}. In that upcoming work, we will also present a comparison of UVLFs at $z \sim 6 - 9$ that we did not show in this paper. \citep[][]{ma-18-a, ma-19} already benchmarked the SMFs/UVLFs from the high-redshift FIRE simulations against those from deep direct observations, and the FIRE-2 values agree well with HST-based SMFs/UVLFs at $z \gtrsim 6$ down to $M_{\rm star} \sim 10^7 \, \rm{M_\odot}$ and $M_{\rm UV} \sim -12$ at $z \approx 6$. The baryonic mass resolution of these high-redshift simulations is good enough ($M_{\rm star} \sim 100 \, \rm{M_\odot}$) to provide SMFs down to $M_{\rm star} \sim 10^3 \, \rm{M_\odot}$. Thus, the FIRE high-redshift suite of simulations provides an excellent theoretical benchmark to the global, unbiased (to the extent to which deep HST fields are unbiased) population of low-mass galaxies against which to compare the SMFs inferred from our suite of MW/LG-like simulations at $z = 0$.

\subsubsection{Limits of resolution}
\label{sec:4.x-resolution-tests}

The FIRE-2 MW/LG-like simulations we use have baryonic mass resolution of $3500 - 7100 \, \rm{M_\odot}$, so we cannot probe galaxies with $M_{\rm star} \lesssim 10^4 \,\rm{M_\odot}$. In upcoming work, we will explore this limit directly, using an ultra-high resolution MW-analogue simulation with baryonic mass resolution of $880 \, \rm{M_\odot}$ (Wetzel et al., in prep.), which will allow us to explore the effects of resolution down $M_{\rm star} \approx 10^3 - 10^{3.5} \, \rm{M_\odot}$ at $z \geq 6$. Additionally, the high resolution ($M_{\rm star} \sim 100\,\rm{M_{\odot}}$) of the high-redshift suite of FIRE simulations that we will be comparing against in upcoming work (Gandhi et al., in prep) will allow us to push this limit as well, as discussed in the previous sub-section~\ref{sec:4.x-lg-representativeness}.

\subsubsection{UV background in FIRE-2}
\label{sec:4.x-uvbackground}

As mentioned in Section~\ref{sec:2.x-fire-sims}, the FIRE-2 simulations include a meta-galactic UV/X-ray background from \citet{faucher-giguere-09} to model the effects of reionisation on cosmic gas. This older model leads to an early timeline of reionisation, with an average neutral hydrogen fraction of $0.5$ by $z \approx 10$, as opposed to the measurements at $z \approx 7.8$ from state-of-the-art observations \citep[such as][]{planck-16}. Additionally, these simulations inadvertently include erroneous heating (due to cosmic rays) of neutral gas at $\leq 1000$ K at $z \gtrsim 10$ \citep{wetzel-23}. At these extremely high redshifts (before reionisation occurs), this spurious cosmic ray heating reduced star formation in low-mass haloes. This is perhaps not as major of an issue, since it acts in the same direction as the too-early reionisation UV/X-ray background, and it does not have an effect after reionisation begins -- so the comparisons we make between $z = 0$ and $z = 6 - 9$ in this paper remain self-consistent.
That being said, to more comprehensively address these concerns, in future work (Gandhi et al., in prep.), we will analyze FIRE-2 simulations re-run without the spurious cosmic ray heating issue, and using an updated model for the UV background \citep{faucher-giguere-20}, which leads to an average reionisation time of $z \approx 7.4$. We will repeat our characterisation of the near-far technique using these simulations; our initial analysis does not show any significant deviations.

\subsection{Discussion}
\label{sec:4-discussion}

We show that by reconstructing star formation histories for galaxies down to $M_{\rm star} \sim 10^{4.5} \, \rm{M_\odot}$ at $z = 0$, the `near-far' reconstruction technique accurately recovers the slope of the low-mass end of the galaxy SMF at $z \sim 6 - 9$ in the proto-LG. Thus, it can provide a powerful complementary approach to direct observations of low-mass galaxies at $z \geq 6$ by probing significantly lower in mass than even the deepest HST/JWST lensing fields.

We showed that the question of mergers and disruption of low-mass galaxies over cosmic time from the proto-LG at $z \gtrsim 6$ to the LG at $z = 0$ does \textit{not} bias the inference of the low-mass slope of the SMF at $z \gtrsim 6$, even if it does only recover $15 - 20$ per cent of the normalisation at all masses we consider. This normalisation issue consists of two effects: low-mass galaxies merging with each other and low-mass galaxies merging into (getting disrupted by) the MW/M31-mass host galaxy. We disentangled these two effects, because we show that $\approx 50$ per cent of the total number of galaxies in the entire progenitor system at $z \approx 7$ end up forming the central MW/M31-mass galaxy, and that this effect is constant with respect to galaxy mass, such that it still does not affect inferences of the slope of the high-redshift SMF. Of the remaining $\approx 50$ per cent of galaxies in the progenitor system at $z \approx 7$, information about $30 - 35$ per cent of them is lost because of mergers between low-mass galaxies before $z = 0$, leaving us with recovery of $15 - 20$ per cent of the overall total number of galaxies in the progenitor system at $z \gtrsim 6$ from the stellar fossil record of surviving low-mass galaxies at $z = 0$. \textit{Most importantly, none of these merger, disruption, and accretion effects appear to bias the inferred low-mass end slope of the reionisation-era SMF in the proto-LG.}

Using the stellar fossil record of low-mass galaxies at $z=0$ in our simulations, we infer an average low-mass end slope of the SMF $\alpha_{\rm cumul} \approx -0.73$ to $-0.98$ from $z \sim 6-9$ (that is, $\alpha_{\rm diff} \approx -1.73$ to $-1.98$). In addition, the average low-mass end slope of the `true' SMF for the proto-MW/LG progenitor systems in our simulations is $\alpha_{\rm diff} \approx -1.68$ to $-2.13$. Comparing this to established values of the low-mass end slope from the literature, we find generally good agreement in most cases. \citet{grazian-15} used deep HST, \textit{Spitzer}, and VLT imaging in the CANDELS-UDS, GOODS-South, and HUDF fields to compute the SMF at $3.5 < z < 7.5$ down to $M_{\rm star} \sim 10^8\,\rm{M_{\odot}}$. They found an average low-mass end slope of $\alpha_{\rm diff} \approx -1.55$ over $5.5 < z < 6.5$, and $\alpha_{\rm diff} \approx -1.88$ over $6.5 < z < 7.5$, which agree broadly with our values. Additionally, we find agreement with the average slopes from \citet{stefanon-21}, who measured SMFs from $z \sim 6 - 10$ using Lyman-break galaxies from the CANDELS and various HST deep fields. They measured average $\alpha_{\rm diff} \approx -1.88$ to $-2.00$ down to $M_{\rm star} \sim 10^7\,\rm{M_{\odot}}$ from $z \sim 6 - 9$.      

Additionally, we find that our median slopes (for both the `true' SMFs of all galaxies in the progenitor systems, as well as the SMFs inferred from the fossil record) are broadly consistent with the inferences of the faint-end UVLF slope made by \citet[][and references therein]{weisz-14c}. They used the actual stellar fossil record from observations of $37$ LG low-mass galaxies, and although their direct inferences are presented only up to $z \sim 5$ (with an average $\alpha_{\rm diff} \sim -1.57$), we find that our slopes agree well with their extrapolations to $z \sim 6 - 9$. Also, \citet{weisz-14c} did not find any signs of a roll-over in the slope down to very faint (low-mass) regimes, and neither do we (to the extent that we can probe given our simulation resolution). Note however, that when compared to low-mass end UVLF slopes from recent JWST observations from CEERS, our slopes at $z \sim 9$ are somewhat shallower relative to those from \citet{leung-23}, who measure an average value of $\alpha_{\rm diff} \approx -2.45$ down to $M_{\rm UV} \approx -17.35$ at $z \sim 9$. A caveat here is that their study presented inferred UVLFs while we only discuss SMFs in their paper, so a direct comparison may not be the most reliable. In an upcoming paper (Gandhi et al., in prep), we will present inferences of the UVLF at $z \sim 6-9$ from our simulations, and make further, more direct comparisons with the established literature.

Thus, we showed that although the near-far technique using low-mass galaxies in the LG is incomplete without reconstructing the entire fossil record of the MW and M31 themselves, this incompleteness has an impact only when considering the normalisation of the inferred SMF at $z \gtrsim 6$. It does not affect the inference of the shape/slope of the SMF, especially at low masses. This is encouraging, because reconstructing the full archaeological record of the MW and M31 is difficult, despite the advent of chemo-dynamical methods to sort through merger and accretion histories \citep[e.g.,][]{cunningham-22, horta-23}. Recently, \citet{brauer-22} showed that when studying the merger/accretion histories of the faintest, lowest-mass galaxies that formed the MW's stellar halo, even chemo-dynamical techniques are likely limited. \textit{We show that applying the near-far technique to only low-mass galaxies in the LG and not to the MW/M31 themselves is an effective approach, and that we can trust the slopes/shapes of the inferred SMF at high redshift.}

Of course, all of this assumes that we can trust the accuracy of CMD-based reconstruction of the SFHs of galaxies today. One potential issue is how well one can get resolved photometry for all stars in the galaxy. JWST is poised to improve upon existing HST measurements in this regard, given its larger aperture and increased range and sensitivity in the infrared \citep{weisz-23}. A second source of uncertainty lies in distinguishing between old stellar populations that are $1 - 2$ Gyr apart in age, in the heterogeneity of the various stellar evolution tracks used for modelling CMDs, and in the inherent uncertainty in stellar evolution models themselves \citep[see][for a comprehensive discussion of the latter]{gallart-05}. JWST, again, will help with the former thanks to its infrared imaging capabilities, because most old stellar populations would be brightest in the infrared. \textit{In Gandhi et al., in prep., we will use synthetic observations of low-mass FIRE-2 galaxies to test the effects of all of these issues on the CMD-based SFH reconstruction method.}

Finally, our analysis in this paper relied on a large spatial selection of low-mass galaxies at $z = 0$, out to $2$ Mpc to the centres of MW- and LG-analogue environments. With the combined power of JWST for obtaining resolved stellar photometry out to larger distances in the LG, and the potential to detect many more extremely faint, low-mass LG galaxies via the Vera Rubin, Euclid, and Nancy Grace Roman observatories, the near-far technique should be easier to apply with a larger sample of LG galaxies in the future. In in an upcoming Gandhi et al., paper, we will test the impact of applying near-far reconstruction out to different distances in the LG (including distance cuts closer to the MW/M31), impact these results. Thus, we hope to make forecasts for which galaxies and regimes these observatories should target in the next $5-7$ years of near-field cosmology.

\section*{Acknowledgements}
\label{sec:acknowledgements}

We thank Jenna Samuel, Isaiah Santistevan, Hyunsu Kong, and Yao-Yuan Mao for valuable discussions that improved this paper. PJG thanks Isabella Trierweiler for help and unwavering support, as always. PJG also thanks Postdog Theodore for the many hours of moral and emotional support, and to the designers of the Star Trek LCARS\footnote{\url{https://memory-alpha.fandom.com/wiki/Library_Computer_Access_and_Retrieval_System}} for inspiring the colour scheme used in all the figures.

This analysis relied on \textsc{NumPy} \citep{numpy-20}, \textsc{SciPy} \citep{scipy-20}, \textsc{AstroPy}\footnote{\url{https://www.astropy.org}}, a community-developed core \textsc{Python} package for Astronomy \citep{astropy-13, astropy-18}, \textsc{Matplotlib}, a \textsc{Python} library for publication-quality graphics \citep{matplotlib-07}, the \textsc{IPython} package \citep{ipython-07}, and the publicly available packages \textsc{GizmoAnalysis} \citep[][available at \url{https://bitbucket.org/awetzel/gizmo\_analysis}]{wetzel-20-gizmo} and \textsc{HaloAnalysis} \citep[][available at \url{https://bitbucket.org/awetzel/halo\_analysis}]{wetzel-20-halo}; as well as NASA's Astrophysics Data System (ADS)\footnote{\url{https://ui.adsabs.harvard.edu}} and the \textsc{arXiv}\footnote{\url{https://www.arxiv.org}} preprint service.

PJG is grateful for support from the Texas Advanced Computing Center (TACC) via a Frontera Computational Science Fellowship. PJG and AW received support from: the NSF via CAREER award AST-2045928 and grant AST-2107772; NASA ATP grants 80NSSC18K1097 and 80NSSC20K0513; HST grants GO-14734, AR-15057, AR-15809, and GO-15902 from STScI; a Scialog Award from the Heising-Simons Foundation; and a Hellman Fellowship. MBK acknowledges support from NSF CAREER award AST-1752913, NSF grants AST-1910346 and AST-2108962, NASA grant 80NSSC22K0827, and HST-AR-15809, HST-GO-15658, HST-GO-15901, HST-GO-15902, HST-AR-16159, HST-GO-16226, HST-GO-16686, HST-AR-17028, and HST-AR-17043 from the Space Telescope Science Institute, which is operated by AURA, Inc., under NASA contract NAS5-26555. RES gratefully acknowledges support from NSF grant AST-2007232 and NASA grant 19-ATP19-0068. AS acknowledges support from HST grants GO-15902, AR-17206 and GO-17216. DRW acknowledges support from NASA through HST grants GO-15901, GO-15902, AR-16159, GO-16273, GO-16226, GO-16686, and JWST-DD-ERS-1334  from the Space Telescope Science Institute, which is operated
by AURA, Inc., under NASA contract NAS5-26555. GS is supported by a CIERA Postdoctoral Fellowship. CAFG received support from NSF through grants AST-2108230, AST-2307327, and CAREER award AST-1652522; from NASA through grants 17-ATP17-0067 and 21-ATP21-0036; from STScI through grant HST-GO-16730.016-A; and from CXO through grant TM2-23005X.

We ran simulations and performed numerical calculations using: the UC Davis computer cluster Peloton, the Caltech computer cluster Wheeler, the Northwestern computer cluster Quest; XSEDE, supported by NSF grant ACI-1548562; Blue Waters, supported by the NSF; Frontera allocations FTA/Hopkins-AST21010 and AST20016, supported by the NSF and TACC; XSEDE allocations TG-AST140023 and TG-AST140064, and NASA HEC allocations SMD-16-7561, SMD-17-1204, and SMD-16-7592; Pleiades, via the NASA HEC program through the NAS Division at Ames Research Center.


\section*{Data Availability}
\label{sec:data-availability}

The \textsc{python} code and data tables used to create each figure are available at \url{https://github.com/pratikgandhi95/nearfar_paper1_protoLG}. The FIRE-2 simulations are publicly available \citep{wetzel-23} at \url{http://flathub.flatironinstitute.org/fire}. Additional FIRE simulation data is available at \url{https://fire.northwestern.edu/data/}. A public version of the \textsc{Gizmo} code is available at \url{http://www.tapir.caltech.edu/~phopkins/Site/GIZMO.html}.



\bibliographystyle{mnras}
\bibliography{near-far-1} 




\appendix

\section{Differential mass functions at \MakeLowercase{z}=0 and \MakeLowercase{z}=7}
\label{appA}

\begin{figure*}
\centering
\begin{tabular}{c c}
\includegraphics[width = 0.45 \linewidth]{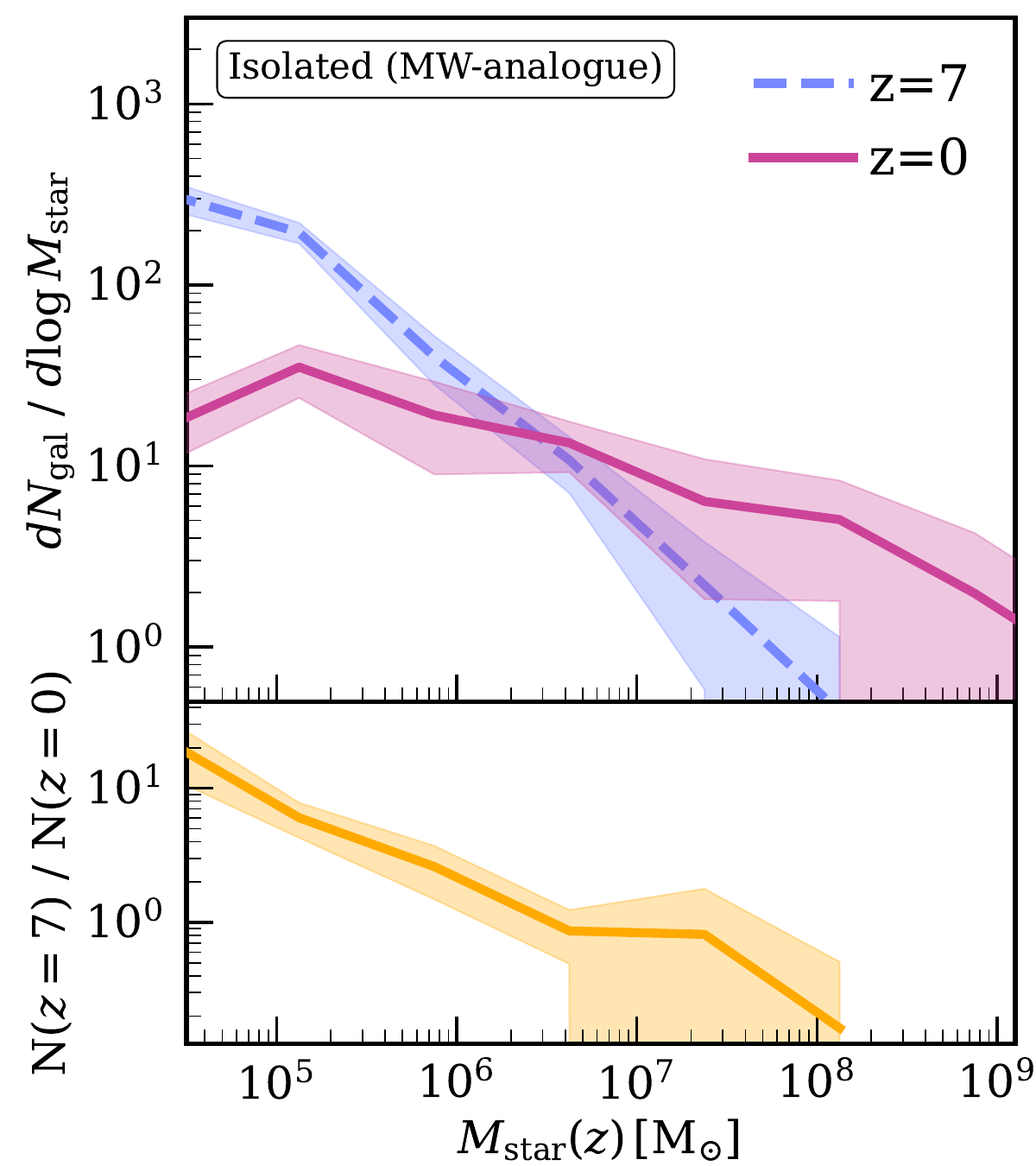}&
\includegraphics[width = 0.45 \linewidth]{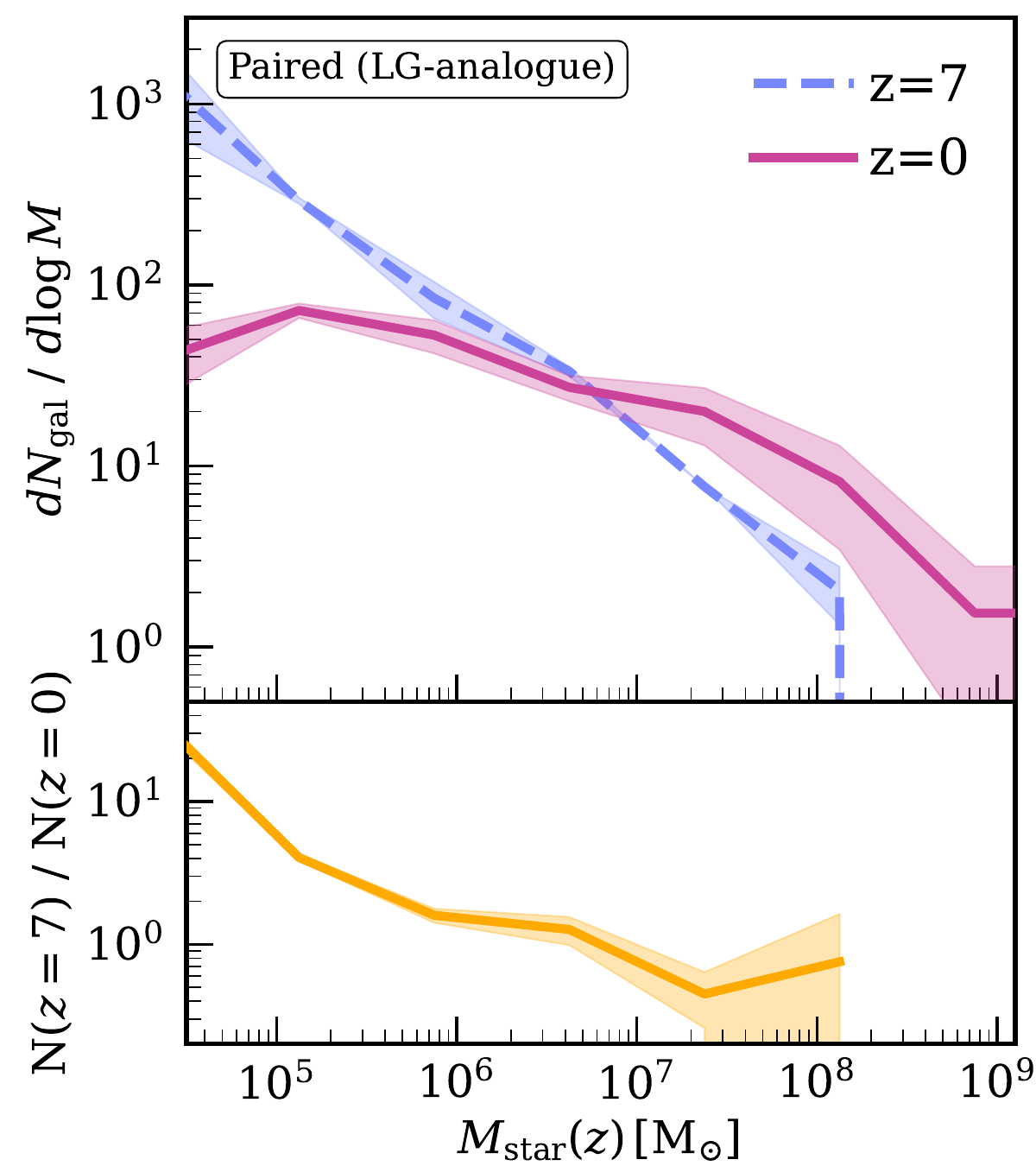}
\end{tabular}
\caption{
\textbf{Same as Figure~\ref{fig:smf-z0-z7}, but for the differential (instead of cumulative) stellar mass function}. Same qualitative takeaways as Figure~\ref{fig:smf-z0-z7}.
}
\label{fig:smf-z0-z7-differential}
\end{figure*}

\begin{figure*}
\centering
\begin{tabular}{c c}
\includegraphics[width = 0.48 \linewidth]{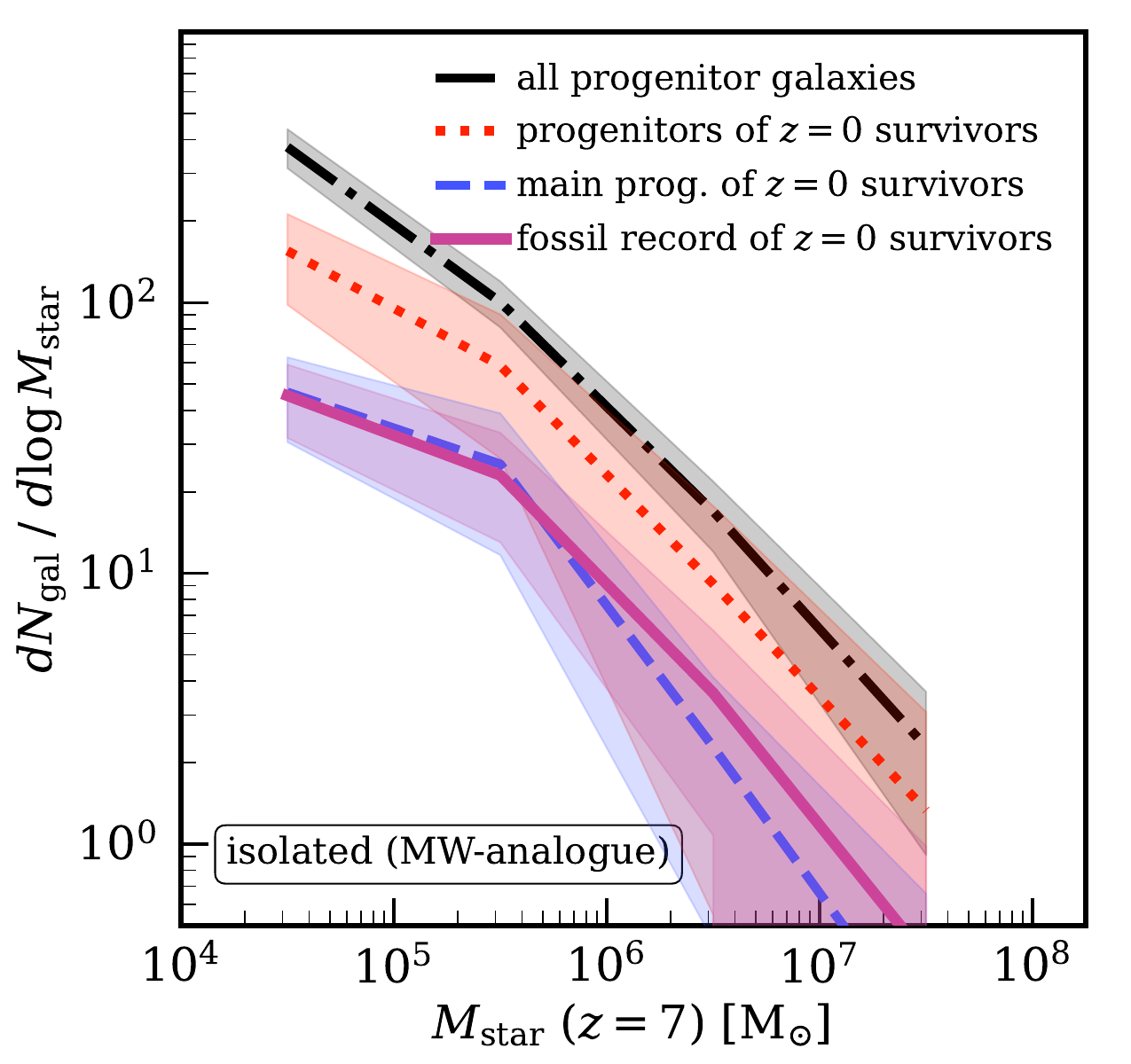}&
\includegraphics[width = 0.48 \linewidth]{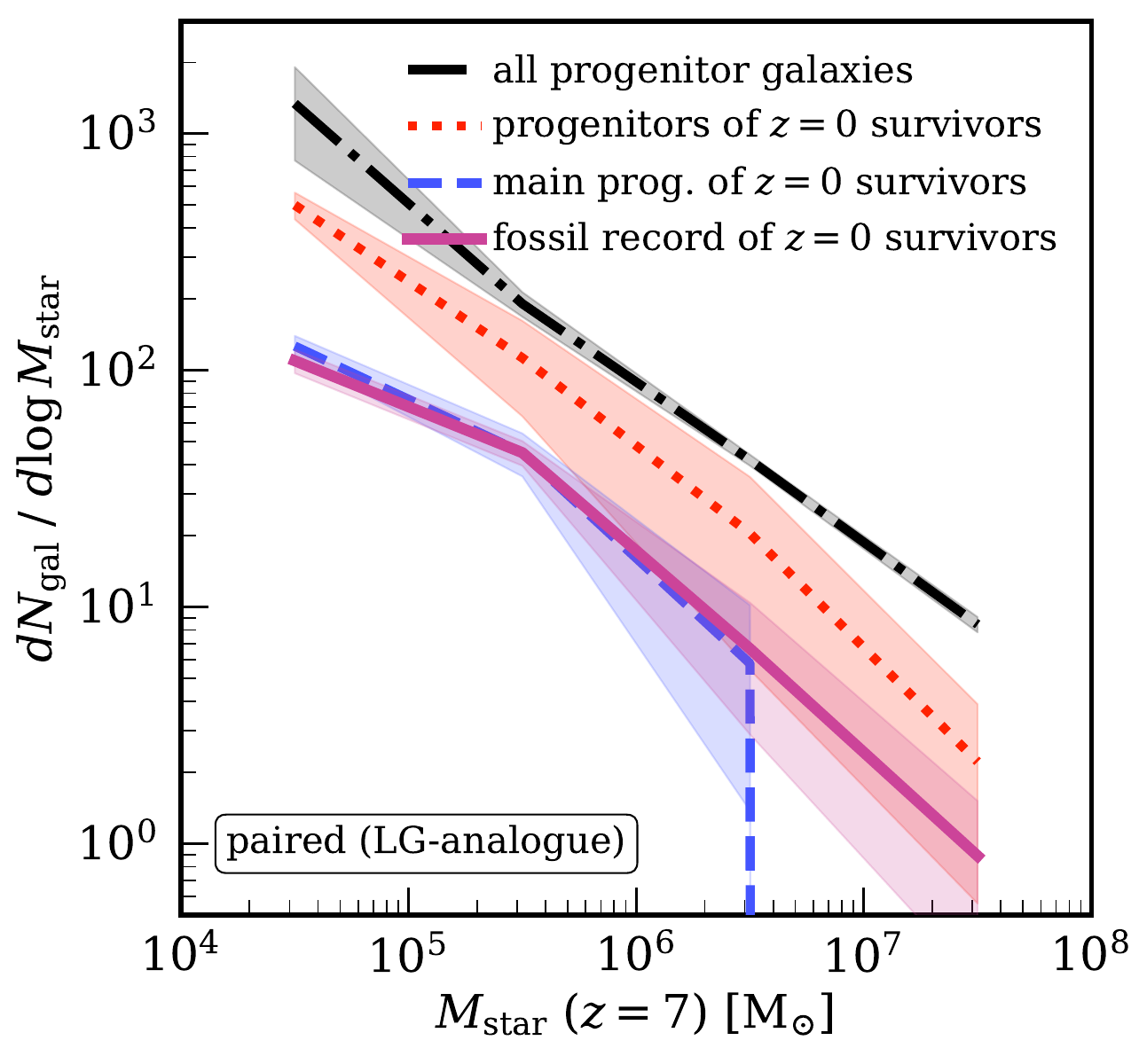}\\
\end{tabular}
\includegraphics[width = 0.75 \linewidth]{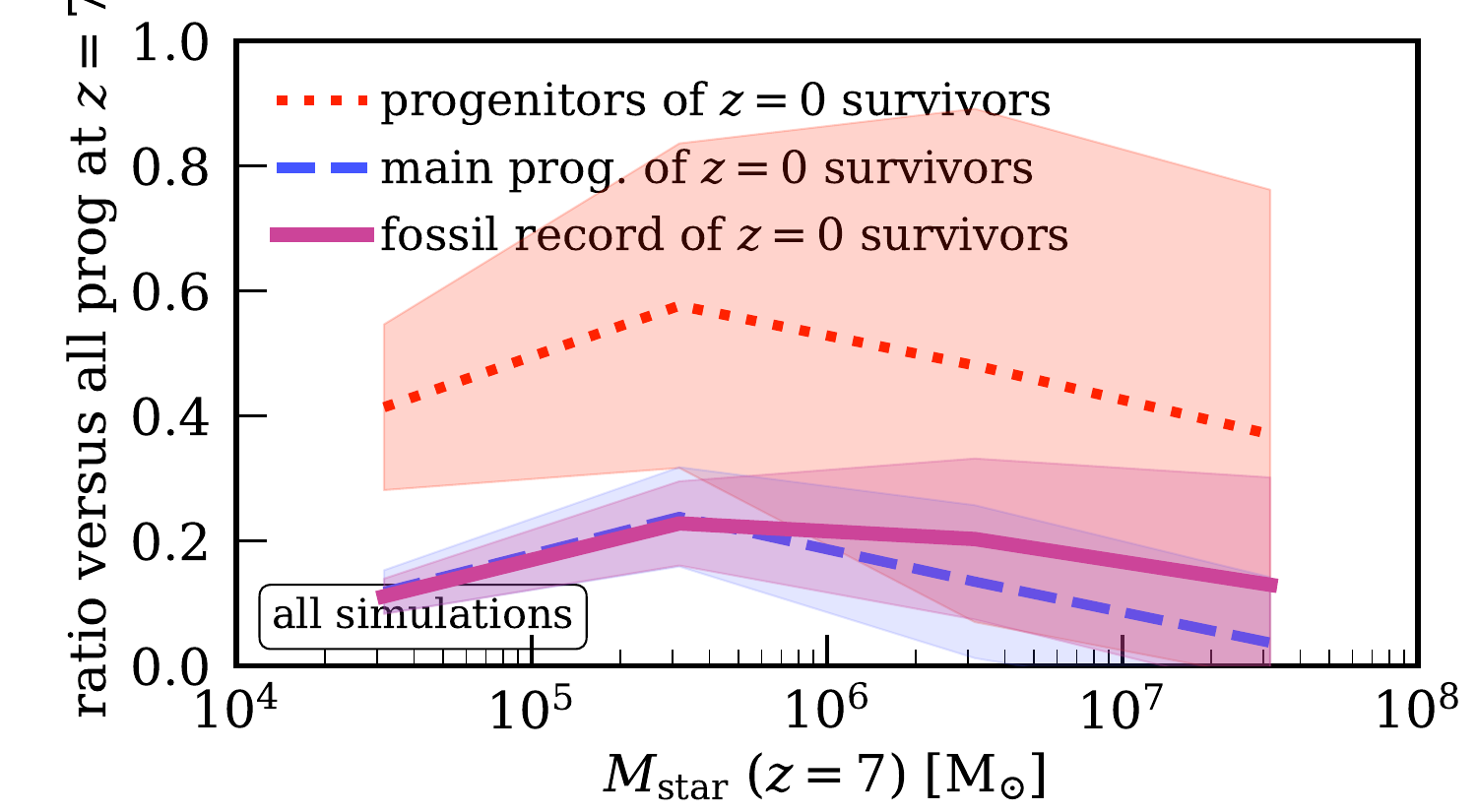}
\caption{
\textbf{Same as Figure~\ref{fig:different-smfs-compared-to-all-at-z7}, but for the differential (instead of cumulative) stellar mass function.} The results are consistent with those shown in Figure ~\ref{fig:different-smfs-compared-to-all-at-z7}.
}
\label{fig:different-smfs-compared-to-all-at-z7-differential}
\end{figure*}

\begin{figure*}
\centering
\begin{tabular}{c}
\includegraphics[width = 0.75 \linewidth]{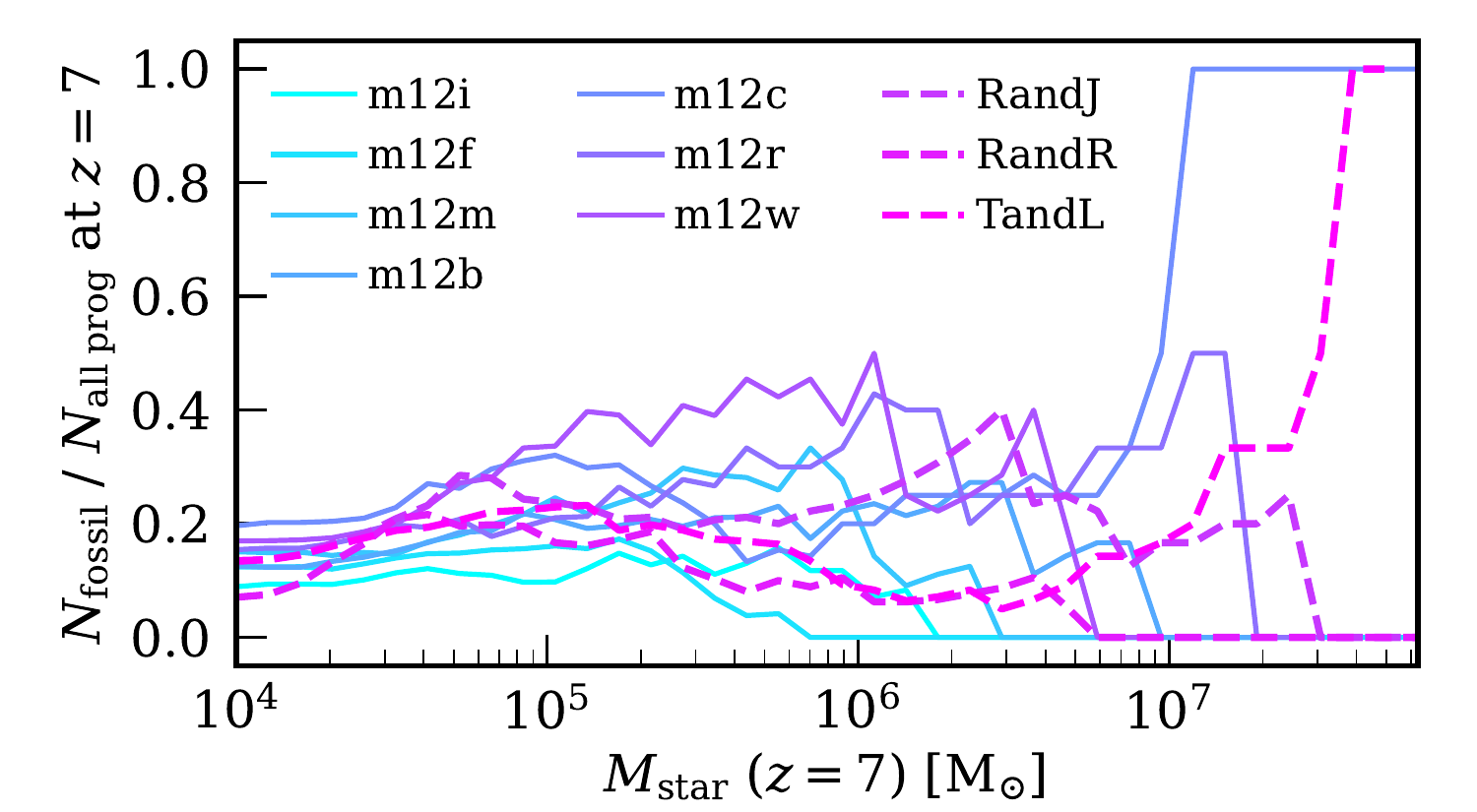} \end{tabular} 
\caption{
\textbf{Same as Figure~\ref{fig:different-smfs-compared-to-all-at-z7} (bottom), for the fossil record at $z = 7$ from surviving low-mass galaxies at $z = 0$, but showing each simulation individually}. 
Although the normalisation at $M_{\rm star} \lesssim 10^7 \, \rm{M_\odot}$ shows some variation, each curve is relatively flat, especially for the LG-like simulations (dashed).
Thus, the flatness of the trend in Figure~\ref{fig:different-smfs-compared-to-all-at-z7} is not simply an artifact of averaging across the simulations, but it reflects the robustness of the `near-far' reconstruction approach applied to an individual system.
}
\label{fig:different-smfs-compared-to-all-at-z7-individual}
\end{figure*}

Figure~\ref{fig:smf-z0-z7-differential} shows the differential versions of the SMF at $z = 0$ and $z = 7$, instead of the cumulative version as in Figure~\ref{fig:smf-z0-z7}. The same key results here are that the steepness of the SMF at $z = 7$ and the lower high-mass cutoff compared to the SMF at $z = 0$ are a natural result of galaxies growing, hierarchically through mergers, over cosmic time.

\section{Differential version of Figure 4}
\label{appB}

Figure~\ref{fig:different-smfs-compared-to-all-at-z7-differential} shows the differential versions of the SMFs, instead of the cumulative versions as in Figure~\ref{fig:different-smfs-compared-to-all-at-z7}. Again, the magenta curve in the bottom panel is roughly constant with stellar mass, implying that the slope of the SMF at $z = 7$ inferred from the stellar fossil record of surviving low-mass galaxies in MW/LG environments at $z = 0$ is accurate, especially at the faint end. We find a similar recovery normalisation of $\approx 10 - 20$ per cent, as in Figure~\ref{fig:different-smfs-compared-to-all-at-z7}.

\section{Individual simulations in Figure 4}
\label{appC}

Figure~\ref{fig:different-smfs-compared-to-all-at-z7-individual} shows the key result from the magenta curve in Figure~\ref{fig:different-smfs-compared-to-all-at-z7} (bottom), except here we show each simulation individually, instead of averaging across them. Especially at low masses ($M_{\rm star} \lesssim 10^7 \, \rm{M_\odot}$), the trends are qualitatively flat for all simulations, implying that our main result, that the slope of the inferred SMF at $z = 7$ is unbiased, still holds. At higher masses, we see divergence for some simulations, but likely because of the small number of more massive galaxies when considering each simulation separately.


\bsp	
\label{lastpage}
\end{document}